\documentclass[letterpaper,journal]{IEEEtran}
\usepackage{amsmath} 
\usepackage{amssymb}  
\usepackage{amsfonts}
\usepackage{multirow}
\usepackage{color}
\usepackage{mathtools}
\usepackage[linesnumbered,ruled,vlined]{algorithm2e}
\usepackage{array}
\usepackage{textcomp}
\usepackage{stfloats}
\usepackage{url}
\usepackage{verbatim}
\usepackage{graphicx}
\usepackage{cite}
\usepackage{float}
\usepackage{url}
\usepackage{physics}
\usepackage{stackengine}
\usepackage{tablefootnote}
\usepackage[utf8]{inputenc}
\usepackage{nomencl}
\usepackage{mathrsfs}
\usepackage{textcomp}
\usepackage[table]{xcolor}
\usepackage{subfigure}
\usepackage{bm}
\usepackage{xtab}
\usepackage{array}

\def\H{\hspace{-1.0mm}}
\def\hH{\hspace{-1.5mm}}
\def\HH{\hspace{-2.0mm}}

\PassOptionsToPackage{caption=false,font=footnotesize}{subfig}
\newtheorem{remark}{Remark}

\allowdisplaybreaks

\begin{document}

% \title{Optimal EV Charging Schedule and Incentivization under Distribution Grid Dynamical Constraints}

% \title{Flexible Scheduling of Large EV Loads for Improving Stability of Power Distribution Systems}
% \title{Co-optimization of Scheduling, Control, and Incentivization for Flexible EV Charging}
\title{Co-Optimization of Scheduling, Control, and Incentivization of Electric Vehicles for Improving Power System Stability}
\author{Amit Kumer Podder, Tomonori Sadamoto, and Aranya Chakrabortty

\thanks{This work was supported in part by the US National Science Foundation under grant ECCS 1931932.}
\thanks{A. Podder and A. Chakrabortty are with the Department of Electrical and Computer Engineering, North Carolina State University, Raleigh, NC 27606. Emails: \tt\small \{apodder, achakra2\}@ncsu.edu}
\thanks{T. Sadamoto is with the Department of Mechanical Engineering and Intelligent Systems, The University of Electro-Communications, Chofugaoka, Chofu, Tokyo. Email: \tt\small sadamoto@uec.ac.jp}
}


\maketitle
%%%%%%%%%%%%%%%%%%%%%%%%%%%%%%%%%%%%%%%%%%%%%%%%%%%%%%%%%%%%%%%%%%%%%%%%%%%%%%%%
\begin{abstract}
We present a finite-horizon, greedy heuristic-based co-optimization and optimal control framework that jointly determines the optimized charging rates, charging times, and an optimal charging control law for electric vehicles (EVs) to improve both voltage stability and small-signal dynamic performance of a power distribution system. The proposed strategy encourages EV drivers to be flexible in choosing their charging rates and times through monetary incentivization under a fixed budget. The design is validated using the IEEE 33-bus distribution system model with ten EV charging stations (EVCSs). Results show that incentivizing drivers to settle for lower charging rates, implemented by a wait-and-save strategy, and rescheduling their charging times strategically can significantly improve the voltage stability index (VSI) of the system as well as the damping factors for the mid-frequency oscillation modes seen in the DC-link bus voltage.

% These results confirm that dynamic, incentive‐driven coordination of DC fast charging can reconcile individual convenience with the stringent stability requirements of modern distribution systems.
\end{abstract}


\noindent {\bf Grid}\\[4pt]
\noindent
\begin{tabular}{p{17mm} p{65mm}}
$n$           & number of buses \\ 
$\mathbb{N}$  & index set of all buses, i.e., $\mathbb{N}:=\{1,\ldots,n\}$. \\
$\mathbb{N}_{\rm P}, \mathbb{N}_{\rm E}, \mathbb{N}_{\rm L}$ & index set of the point of common coupling (PCC) bus, EVCS buses, and non-EV load buses. These sets are disjoint, and $\mathbb N = \mathbb N_{\rm P}\cup \mathbb N_{\rm E}\cup \mathbb N_{\rm L}$.  \\
$\mathbb N_{\rm rec}, \mathbb N_{\rm send}$ & index set of receiving/sending bus \\
$\mathcal N_k \subset \mathbb N_{\rm rec}$ & index set of receiving buses to the $k$-th bus for $k \in \mathbb N_{\rm send}$. \\
$\mu_{v}^{(\tau)}, \mu_{v}^{p(\tau)}$ & a binary variable that equals to $1$ if EV $v$ is assigned to time slot $\tau$, otherwise $0$. $\mu_{v}^{p(\tau)}$ is the preferred value declared by EV $v$. \\
$V_k \in \mathbb R$ & voltage magnitude at the $k$-th bus, for $k \in \mathbb N$. \\
$P_k^L$, $Q_k^L$ & constant real/reactive power consumed by non-EV load at the $k$-th bus, for $k \in \mathbb N_{\rm L}$. \\
${\rm VSI}_h$ & Voltage stability index at the $h$-th bus, $h\in \mathbb N_{\rm rec}$. \\
${\rm VSI}, {\rm VSI}_{\rm min}$ & ${\rm VSI} := [{\rm VSI}_h]_{h \in \mathbb N_{\rm rec}}^{\top}$, \&  minimum of ${\rm VSI}$  \\
% ${\rm VSI}_{\rm min}$ & the minimum entry of ${\rm VSI}$.\\
\end{tabular}

\vspace{3pt}
\noindent {\bf EV}\\[4pt]
\noindent
\begin{tabular}{p{17mm} p{65mm}}
$N_{\rm V}$ & number of EVs  \\
$\mathcal V$ & index set of EVs, i.e., $\mathcal V:=\{1,\ldots, N_{\rm V}\}$.   \\
$v \in \mathcal V$ & index of EV \\
$i_v^{\min}, i_v^{\max}$ & min/max charging current for EV $v$ \\
$i^{\star(\tau)}_v$ & optimal charging current for EV $v$ at slot $\tau$ \\
\end{tabular}

\vspace{2pt}
\noindent {\bf EVCS}\\[4pt]
\noindent
\begin{tabular}{p{17mm} p{65mm}}
$N_{\rm E}$ & number of EVCSs \\
$\mathcal E$ & index set of EVCSs, i.e., $\mathcal E:=\{1,\ldots, N_{\rm E}\}$. Each EVCS is assumed to be connected to a single, distinct bus without overlap \\
$e \in \mathcal E$ & index of EVCS  \\ 
$e_v^p$ & preferred choice of EVCS for the driver of EV $v\in\mathcal{V}$\\
$v_e^{dc*}$ & DC link voltage at EVCS $e$ \\
$P_e^{\rm max}$ & maximum charging power at EVCS $e$ \\
$P^{\star(\tau)}_e$ & optimal charging power at EVCS $e$ at slot $\tau$ \\
$\pi_{e}$ & per-kWh energy price at EVCS $e$ 
\end{tabular}

\vspace{6pt}
\noindent {\bf Time slot}\\[4pt]
\noindent
\begin{tabular}{p{17mm} p{65mm}}
$N_{\rm T}$ & number of time slots \\
$\mathcal T$ & index set of time slots, i.e., $\mathcal T:=\{1,\ldots, N_{\rm T}\}$. \\
% The slot duration length is assumed to be common among EVCSs.  \\
$\tau \in \mathcal T$ & index of time slot \\
$\tau_v^p$ & preferred time slot of the driver of EV $v\in\mathcal{V}$\\
\end{tabular}

\vspace{6pt}
\noindent {\bf Charging demand and incentivization}\\[4pt]
\noindent
\begin{tabular}{p{17mm} p{65mm}}
$T^{\rm min}_{v},T^{\star}_{v}$ & charging duration at the maximum and optimal charging power rate for EV $v$\\
$M_v^{\rm C\star}, M_v^{\rm S\star}$ & dollar incentive caused by charging current changes and by time shift changes \\
$C^{D}_{v},C^{\rm net}_{v}$ & demanded charging cost and the net amount including incentive
\end{tabular}
\vspace{4pt}

% \makenomenclature
% \nomenclature{\(P^{eD},i^{eD}\)}{demanded charging power and current vector}
% \nomenclature{\(P^{e*},i^{e*}\)}{optimal charging power and current vector}
% \nomenclature{\(E^{D}\)}{energy demand vector from EVs}
% \nomenclature{\(\mathcal{C}^{tD},\mathcal{C}^{t\ast}\)}{charging/discharging duration at the demanded and optimal charging/discharging power rate}
% \nomenclature{\(\mathcal{C}^{\mathcal{P}D},\mathcal{C}^{\mathcal{P}\ast}\)}{charging/discharging price at the demanded and optimal charging/discharging power rate}
% \nomenclature{\(\mathcal{W}^{t},\mathcal{I}^e\)}{waiting time and dollar incentives}
% \printnomenclature

{\it Mathematical Notations:} 
% We denote the sets of real and complex numbers by $\mathbb{R}$ and $\mathbb{C}$, respectively, and 
We denote the set of non-negative (resp. positive) real numbers as $\mathbb R_{\geq 0}$ (resp. $\mathbb R_{> 0}$). ${\bf 1}_n := [1, \ldots, 1]^{\top} \in \mathbb R^n$. Given $x_{k_1}, \ldots, x_{k_w} \in \mathbb R^{1\times N}$ and $\mathcal K := \{k_1, \ldots, k_w\}$, the quantity $[x_k]_{k \in \mathcal K} := [x_{k_1}, \ldots, x_{k_w}] \in \mathbb R^{1 \times wN}$. For $\mathbb N := \{1, \ldots, n\}$, we denote the block-diagonal matrix with matrices $M_1, \cdots, M_n$ on its diagonal blocks as ${\rm diag}(M_k)_{k \in \mathbb N}$. The symbol $\otimes$ denotes the Kronecker product. For $x := [x_1, \ldots, x_n]^{\top} \in \mathbb R^n$ and $\beta_1, \ldots, \beta_n \in \mathbb R$, we define $\|x\|_{2,\beta} := x^{\top}{\rm diag}(\beta_k)_{k \in \{1,\ldots,n\}}x$ and $\|x\|_{\infty,\beta} := \max\{\beta_1x_1, \ldots, \beta_nx_n\}$, where $\|x\|_{2} := x^{\top}x$. 
Given $X \geq 0$, we denote its square root $X^{\frac{1}{2}} \geq 0$ such that $X=X^{\frac{1}{2}}X^{\frac{1}{2}}$. Given a stable strictly proper transfer matrix $G(s)$, its $\mathcal H_2$-norm is defined as $\|G(s)\|_{\mathcal H_2} := \left(\frac{1}{2\pi}\int_{-\infty}^{\infty}{\rm tr}(G(j\omega)G^{\top}(j\omega))d\omega\right)^{\frac{1}{2}}$ where ${\rm tr}$ is the trace operator. For any symbol $\bullet$, we denote its setpoint as $\bullet^{*}$.
% For three vectors $x_1$, $x_2$, $x_3$ of same dimension, $x_1\in[x_2,x_3]$ means each element of $x_1$ is lower and upper bounded by the corresponding elements of $x_2$ and $x_3$.
%%%%%%%%%%%%%%%%%%%%%%%%%%%%%%%%%%%%%%%%%%%%%%%%%%%%%%%%%%%%%%%%%%%%%%%%%%%%%%%
\section{Introduction}\label{sec_intro}
% \subsection{Motivation}

%start directly with EV trucks - red text. Say how much to charge and when to charge can have huge impact...the how much part was addressed in prior paper...the two are not decoupled problems, so in this paper we show that not only choosing optimal charging rate but also choosing optimal charging time is important... the price paid is incentivization - first via wait and save, and second for time shifting once the user fixes its preferred charging station. we formulate a co-optimization problem - and show how the DSO can maintain the balance between grid stability and the price expenditures. the stability is characterized in terms of small-signal stability where H2-norm of the TF is minimized - this is shown to improve the small-signal damping of eignevalues with frequency ranges of *** in the state repsonses, especially the dc-link bus voltage which is an important variable of interest as it determines the quality of charging.  \\

\IEEEPARstart{I}ncreased adoption of electric vehicles (EVs), especially heavy-duty EVs such as electric trucks and electric buses, has led to the emergence of diverse charging solutions, including medium-voltage AC chargers in parking lots and DC fast chargers at public charging stations \cite{ev1}. As the number of EVs scales and becomes an integral part of highway traffic, one challenge that distribution system operators are unanimously concerned about is on how to schedule, control, and coordinate the charging of several hundred EVs within a short period of time to avoid unexpected peak load pressure that can easily cause fluctuations in voltage and frequency at critical buses in the system. Multiple papers have been written on optimal EV charging and scheduling, such as \cite{ev3,Ref_3,Ref_4,ev4,ref_5}, but mostly aimed at steady-state operations such as load flattening, voltage and frequency regulation, electricity pricing, and modeling of the equilibrium behavior of EV queues. Fair allocation of benefits between EV charging station (EVCS) owners and EV customers has been addressed in \cite{benefit_allocation} using game theory, and monetary incentivization of EV customers to encourage flexible charging has been studied in \cite{bi_level}, but, again, from a steady-state or quasi-steady-state point of view. Therefore, most of the existing optimal charging scheduling solutions are carried out in AC distribution networks by solving static convex or non-convex optimization problems without accounting for the dynamics coming from the power-electronics-based EVCS side. Moreover, these methods primarily target slow and medium-speed chargers by considering them as aggregated passive loads, making them unsuitable for emerging AC/DC distribution networks where DC fast charging stations, interfaced via voltage source converters and governed by fast-acting local control logic, are integrated into the power distribution grid.
This raises a critical question: how can we design coordinated charging strategies that not only account for the dynamic behavior of high-power DC fast charging stations but also ensure better steady-state voltage stability index (VSI) within the distribution grid?

To address this gap, in our recent paper \cite{tcst}, we proposed a co-design method where a system operator can dictate the {\it optimal} charging rate for an EV (that may be less than the demanded rate) and design an optimal charging control law using state-feedback linear-quadratic regulator (LQR) control such that the combined effect of the two significantly improves the small-signal dynamic performance and VSI of the distribution network. The improvement came at the cost of incentivizing EV drivers to wait an additional amount of time in the charging queue, referred to as {\it wait-and-save}. The strategy was shown to be highly effective in damping the fast and medium-frequency oscillation modes seen in the temporal responses of the DC-link voltages at the charging stations following sudden plug-ins and plug-outs of the EV loads. The co-design was developed, however, assuming an infinite time horizon for the LQR, meaning that vehicles were assumed to arrive at EVCSs randomly over time without following any predetermined schedule.
% \red{Physically, the impact of EV charging depends not only on the {\it charging rate} but also on {\it when} the EVs charge. Shifting charging times can redistribute EV demand across EVCSs and reduce unfavorable loading at electrically weak buses, thereby improving the overall VSI and dynamic performance. This motivates the incorporation of coordinated time-slot scheduling into the co-design framework of \cite{tcst}.}

% \red{As the dynamic response of the grid state variables as well as the voltage stability margin will depend not only on the {\ charging rate} of the vehicles but also on {\it when} they charge, adding a co-ordinated time-scheduling objective to this co-design framework can improve the closed-loop dynamics of the grid even further.} 

% Accordingly, in this paper, we present a finite-horizon, greedy heuristic-based co-optimization and optimal control framework that jointly determines the optimal charging rates, the optimal charging times, and an
% optimal LQR law for EVs to improve both stability and dynamic performance metrics for the grid by encouraging EV drivers to be flexible in choosing their charging rates and times through monetary incentivization.
In this paper, we study how the additional flexibility of {\it scheduling} EVs for charging at different charging stations at different times of the day can help in improving grid dynamics even further. We present a finite-horizon, greedy heuristic-based co-optimization and optimal control framework that jointly determines the charging rates, charging times, and a LQR control law to improve voltage stability and small-signal dynamic performance by encouraging EV drivers to be flexible in choosing their charging rates and times in exchange for monetary incentives. Unlike \cite{tcst}, however, the incentivization now must compensate for not only {\it wait-and-save} but also {\it time-shifting} when the operator decides to change the time of charging for any EV driver from their preferred option.  The optimization variables include the charging rate setpoint, the LQR controller, and a binary variable that decides if a given EV should charge at a given time slot or otherwise. The resulting problem is solved in two stages: the first stage optimizes the charging rates and LQR controllers for the preferred charging times, while the second stage greedily explores feasible time-slot reassignments and re-optimizes the affected variables. 
% \red{The resulting problem is solved in two stages: the first stage optimizes the charging rates and LQR controllers for the preferred charging times, while the second stage greedily explores feasible time-slot reassignments and re-optimizes the affected variables. Finally, to account for user behavioral uncertainty, we evaluate different incentive-option selections by EV drivers and quantify their resulting impacts on the grid's stability and dynamic performance.}

% Multiple incentive options are proposed for enabling real-time participation of EV drivers. 

In summary, the two primary contributions of this work are: 
\begin{enumerate}
    % \item We develop a combinatorial co-optimization and optimal control algorithm for EV charging that optimizes both small-signal dynamic performance and steady -state VSI of a distribution grid while evaluating the right incentives for flexible participation of EV users. 
   \item We develop a finite-horizon, greedy-heuristic-based co-optimization and control algorithm for EV charging that optimizes both small-signal dynamic performance and steady-state VSI of a distribution grid while evaluating the right incentives for flexible participation of EV users. 
   % \item We propose a deployment strategy for this algorithm using multiple incentive options considering user uncertainty in user selection that can help EV users in determining the best possible ways to engage with the proposed scheme.
   \item We propose a deployment strategy with multiple incentive options that accounts for uncertainty in EV user selection and evaluates its impact on system performance while enabling users the best possible ways to engage with the proposed scheme.
\end{enumerate}
We demonstrate our algorithm using a variant of the IEEE 33-bus radial feeder distribution system model with multiple EVCSs distributed across different buses.  
% \subsection{Organization}
% \IEEEPARstart{A} 
% The algorithm is also extended to demonstrate the trade-off between price, small-signal stability, and voltage stability of the grid when the EVCSs operate in bidirectional charging mode. Results are validated using numerical simulations of the modified IEEE 33-bus radial distribution system model with multiple EVCSs. The benefits and drawbacks of the proposed controller on the closed-loop time response of the grid are reported. 
The rest of the paper is organized as follows. Section II recapitulates the nonlinear state-space model of the EVCS-integrated distribution grid. Sections III and IV present the problem formulation and solution strategy of the proposed approach, followed by numerical results in Section V. Section VI concludes the paper.

\section{Model of EVCS-integrated power system}\label{sec_model}
%\subsection{Model of the EVCS-Integrated Power System}\label{sec_model_sub}
%We consider an EVCS-EV integrated power distribution grid, where our goal is to determine the optimal charging power rate and the optimal scheduling of EV charging using an optimal feedback controller, while maintaining a balance between system stability and EV user convenience. For simplifying the modeling, we assume that the grid is a radial power network. The assumption does not impact our problem formulations or strategies. 

Consider a power distribution system with $n$ buses and $p$ EVCSs. 
We assume that each EVCS has a fixed capacity and only operates in grid-to-vehicle (G2V) mode. Each EVCS consists of a centralized AC/DC converter with a front-end LCL filter for creating a common DC bus, and  DC/DC converters, which act as the charging ports. For more details about the model, please see \cite{tcst, LCL_design}. 
The synchronization between the EVCS and the network is ensured by a phase-locked loop (PLL), which provides the synchronous phase angle required for designing the controller of the converter.
Each EVCS may have multiple charging ports connected to its respective DC bus, where every charging port has similar internal dynamics \cite{eig_paper_5}. As shown in \cite{tcst}, the dynamics of the charging ports are ignored for simplicity, as the time-scale of those dynamics is significantly faster than that of the dominant oscillations exhibited in typical EV plug-in or plug-out responses of the DC-link voltage of the EVCS. Short-line models are considered for tie-lines between consecutive buses.

The operating point of the distribution system model depends on which EV is connected to which EVCS, and also on the charging current used for charging the EVs at each EVCS. Each EV is assumed to charge at a single pre-specified EVCS (which is connected to a fixed bus) at a time. We denote the {\it preferred EVCS} for the $v^{th}$ EV as $e_v^p$ for $v \in \mathcal{V}$. If multiple EVs share the same EVCS, they cannot charge at the same time. Accordingly, we partition time into multiple {\it time-slots} (e.g., time-slot 1 can mean 9 AM-10 AM), and schedule the time at which each EV charges at its preferred EVCS under the constraint that each EVCS can serve at most one EV per time-slot. To this end, for each EV $v \in \mathcal{V}$ and time-slot $\tau \in \mathcal{T}$, we define a binary variable $\mu_{v}^{(\tau)}$ indicating whether the EV is charging at the preferred EVCS or not. In other words, $\mu_{v}^{(\tau)} = 1$ means that EV $v$ connects to its preferred EVCS $e_v^p$ at time slot $\tau$. Let the charging current of this EVCS at that time-slot be denoted as $i^{*(\tau)}_{v}$. We define the following two vectors for $\tau \in \mathcal T$:
\begin{align}
 \mu^{(\tau)} &:= [\mu_{1}^{(\tau)}, \ldots, \mu_{N_{\rm V}}^{(\tau)}]^{\top} \in \{0,1\}^{N_{\rm V}} \label{def_mu} \\
 i^{*(\tau)} &:= [i_{1}^{*(\tau)}, \ldots, i_{N_{\rm V}}^{*(\tau)}]^{\top} \in \mathbb R^{N_{\rm V}}. \label{def_iestar} 
\end{align}
%\begin{align}
% \mu &:= [\mu_{1,1}, \ldots, \mu_{1,N_{\rm T}}, \mu_{2,1}, \ldots, \mu_{N_{\rm V},N_{\rm T}}]^{\top} \in \%{0,1\}^{N_{\rm V}N_{\rm T}} \label{def_mu} \\
% i^{*} &:= [i_{1,1}^{*}, \ldots, i_{1,N_{\rm T}}^{*}, i_{2,1}^{*}, \ldots, i_{N_{\rm V},N_{\rm T}}^{*}]^{\top} \in \mathbb R^{N_{\rm V}N_{\rm T}}. \label{def_iestar} 
%\end{align}
The state-space model of the power system can then be written in a compact form by the following differential-algebraic equations (DAEs)
\begin{equation}\label{overall_ss}
%   \sum_k:
   \Sigma^{(\tau)}:\begin{cases}
   \dot{x}^{(\tau)}=f(x^{(\tau)},y^{(\tau)},u^{(\tau)};~i^{*(\tau)}, \mu^{(\tau)})\\
   0=g(x^{(\tau)},y^{(\tau)};~i^{*(\tau)}, \mu^{(\tau)}),\\
\end{cases}
\end{equation}
where, 
\begin{equation} \label{eq1}\nonumber
\begin{split}
 x^{(\tau)} &:=\hspace{-1mm} [[(x_{e}^{(\tau)})^\top]_{e \in \mathcal E}^{\top}, [i_{kh}^{d(\tau)},i_{kh}^{q(\tau)}]^{\top}_{k \in \mathbb N_{\rm send},h\in\mathcal{N}_k}]^{\top}\hspace{-2mm} \in \hspace{-1mm}\mathbb R^{n_x}\\
 x_{e}^{(\tau)}  &:=\hspace{-0.5mm}  [\delta_{e}^{(\tau)},\zeta_{e}^{(\tau)}, i_{e}^{gd(\tau)}, i_{e}^{gq(\tau)}, i_{e}^{cd(\tau)}, i_{e}^{cq(\tau)}, v_{e}^{cd(\tau)}, v_{e}^{cq(\tau)}, \\
 &\hspace{30mm} \Psi_{e}^{(\tau)},\chi_{e}^{d(\tau)},\chi_{e}^{q(\tau)}, v_{e}^{dc(\tau)}]^{\top}\hspace{-1mm} \in\mathbb{R}^{12} \\
u^{(\tau)}&:=\hspace{-0.5mm}\left[\Delta m_{e}^{d(\tau)},\Delta m_{e}^{q(\tau)},\Delta i_{e}^{(\tau)}\right]^{\top}_{e \in \mathcal E}\in\mathbb{R}^{3N_{\rm E}}\\
y^{(\tau)}&:=\hspace{-0.5mm}\left[V_{k}^{d(\tau)},V_{k}^{q(\tau)}\right]^{\top}_{k \in \mathbb N}\in\mathbb{R}^{2n},
\end{split}
\end{equation}
with $n_x := 12N_{\rm E}+2|\mathbb N_{\rm send}|$. Here, $x_e^{(\tau)}$ denotes the state vector for the EVCS that is indexed as $e$ (simply referred to as EVCS $e$). The state vector consists of states from the phase-locked loop $\delta_{e}^{(\tau)},\zeta_{e}^{(\tau)}$, from the front-end LCL filter $i_{e}^{gd(\tau)}, i_{e}^{gq(\tau)}, i_{e}^{cd(\tau)}, i_{e}^{cq(\tau)}, v_{e}^{cd(\tau)}, v_{e}^{cq(\tau)}$, states of the inner-loop and outer-loop current controller of the AC/DC converter $\Psi_{e}^{(\tau)},\chi_{e}^{d(\tau)},\chi_{e}^{q(\tau)}$, and the DC-link voltage $v_{e}^{dc(\tau)}$. The currents $i_{kh}^{d(\tau)}$ and $i_{kh}^{q(\tau)}$ are the $d$-$q$ axis branch currents of the network. $V_k^{d(\tau)}$ and $ V_k^{d(\tau)}$ are $d$-and $q$-axis components of the bus voltage $V_k$ for $k\in \mathbb N$. Finally, $\Delta m_{e}^{d(\tau)}$, $\Delta m_{e}^{q(\tau)}$, and $\Delta i_e^{(\tau)}$ are the controllable modulation indices and charging current inputs, respectively, that will be used as control inputs for our design. The exact definitions of the nonlinear functions $f(\cdot)$ and $g(\cdot)$ follow respectively from the converter model and the power flow equations, detailed expressions for which can be founded in \cite{tcst}, and are skipped here for brevity. % \footnote{Note that, whereas all state variables in \cite{acc} are defined per bus, in this paper the EVCS state variables are defined per EVCS.}. 

\vspace{-0.15in}

\subsection{Linearized Model}
To simplify the subsequent design process, we consider a reduced-order linearized model of \eqref{overall_ss} around $i^{*(\tau)}$ in \eqref{def_mu}-\eqref{def_iestar} based on a singular-perturbation-based model reduction that extracts the dominant (i.e., slow) dynamics from the linearized model of $\Sigma^{(\tau)}$ \cite{khalil_book}. The operating point $\{x^{*(\tau)}, u^{*(\tau)}\}$ depends on the choice of $\{i^{*(\tau)}, \mu^{(\tau)}\}$. 
The algebraic variable $y^{(\tau)}$ in \eqref{overall_ss} is eliminated, and the model is then linearized around the equilibrium point $\{x^{*(\tau)},\,u^{*(\tau)}\}$ by Jacobian linearization to obtain a small-signal state-variable model, where, the control input $u^{(\tau)}$ is already in the small-signal form, i.e., $u^{*(\tau)} = 0$.
% \red{Note that the sensitivity of the resulting stability performance to model mismatch and uncertainty was also evaluated in \cite{tcst}, where the model was shown to exhibit low sensitivity to such uncertainties. This prior validation supports the use of the linearized model for the stability-oriented optimization considered in this work.} 
For a given $\{i^{*(\tau)}, \mu^{(\tau)}\}$, the linearized model of \eqref{overall_ss} with the performance output $\hat{z}^{(\tau)}$ is uniquely determined as 
\begin{equation}\label{Transfer_func}
    G_{i^{*(\tau)},\mu^{(\tau)}}:~\begin{cases}
         \dot{\hat{x}}^{(\tau)}=A(i^{*(\tau)},\mu^{(\tau)})\hat{x}^{(\tau)} + B(i^{*(\tau)},\mu^{(\tau)})u^{(\tau)}, \\
        \hat{z}^{(\tau)} = Q\hat{x}^{(\tau)} + R u^{(\tau)},
    \end{cases}
\end{equation}
with $\hat{x}^{(\tau)}(0)=\hat{x}^{(\tau)}_0$ where, 
\begin{equation}
 \hH \hat{x}^{(\tau)}:={\mathscr P}^{(\tau)}(x^{(\tau)}-x^{*(\tau)}),
 Q := [Q_{\hat{x}}^{\frac{1}{2}}, 0]^{\top}\hspace{-2mm}, R := [0, R_{u}^{\frac{1}{2}}]^{\top}, \label{defxhat}
\end{equation}
and $\hat{z}^{(\tau)}$ represents an output signal of interest that can help the operator in evaluating the damping performance of the small-signal model. 
$\hat{x}^{(\tau)} \in \mathbb  R^{7N_{\rm E}}$ is a reduced-dimensional state vector that consists of only the physical state variables of the EVCSs, and excludes the five non-physical states per EVCS, namely,  $(\delta_{e}^{(\tau)},\zeta_{e}^{(\tau)})$ coming from the PLL, and $(\Psi_{e}^{(\tau)},\,\chi_{e}^{d(\tau)},\,\chi_{e}^{q(\tau)})$ coming from the internal PI controllers of the converter, as well as the two line states $(i_{kh}^{d(\tau)},\,i_{kh}^{q(\tau)})$ per line. This dimensionality reduction phenomenon is detailed in \cite{tcst}. An equivalent interpretation is that the time constants associated with these states are remarkably faster than those with the physical states of the EVCSs, indicating that they can be eliminated using a time-scale separation. Elimination of these unimportant states facilitates both computation and communication needed for the state feedback controller that will be designed in the forthcoming sections. The matrix ${\mathscr P}^{(\tau)} \in \mathbb R^{7N_{\rm E} \times (12N_{\rm E}+2|\mathbb N_{\rm send}|)}$ is constructed accordingly using singular perturbation theory \cite{khalil_book}, resulting in $\hat{x}^{(\tau)}:=[i_{e}^{gd(\tau)},\, i_{e}^{gq(\tau)}, \,i_{e}^{cd(\tau)}, \,i_{e}^{cq(\tau)}, \,v_{e}^{cd(\tau)},\, v_{e}^{cq(\tau)}, \, v_e^{dc(\tau)}]^{\top}$ for $e \in \mathcal E$. The matrices $Q_{\hat{x}}\geq 0\in\mathbb{R}^{7N_{\rm E}\times7N_{\rm E}}$, and $R_u > 0\in\mathbb{R}^{3N_{\rm E}\times3N_{\rm E}}$ are constant weight matrices that can be designed by the system operator.

\vspace{-0.15in}
\subsection{VSI}
We close this section by recalling the definition of VSI that will be used in the problem formulation in the next section. The VSI represents the voltage stability margin at any given bus of the distribution network and is defined at each receiving end bus. 
% \red{Importantly, the VSI explicitly captures the effect of EV charging power on the voltage-stability margin through the network voltage, line impedance, and loading conditions.} 
For a radial network, every receiving end bus only has one sending end bus.%(please note that the reverse may not be true, as the sending end bus may provide power to multiple receiving end buses). 
Since the EVCS to which EV $v$ connects is fixed, for each EVCS $e$, the identity of the EV that connects to EVCS $e$ at time slot $\tau$ is uniquely determined. We denote this EVCS as $v_e^{p(\tau)} \in \mathcal V$ \footnote{In contrast to $e_v^p$, $v_e^{p(\tau)}$ depends on $\tau$ because multiple EVs may prefer the same EVCS at different time slots.}. Moreover, the bus to which EVCS $e$ is connected is also unique; we denote this bus as $k_e \in \mathbb N_{\rm E}$. The steady-state charging power at EVCS $e$ at time slot $\tau$ is then defined as 
\begin{align}
    P_{k_e}^{{\rm Ch}*(\tau)} := \mu_{v_e^{p(\tau)}}^{(\tau)}i_{v_e^{p(\tau)}}^{*(\tau)}v_{e}^{dc*(\tau)}, \quad e \in \mathcal E, 
\end{align}
where, $v_{e}^{dc*}$ is the steady-state DC-link voltage of the EVCS $e$.  
Based on the above definition, for $h \in \mathbb N_{\rm rec}$ and $k_{h} \in \mathbb N_{\rm send}$, following \cite{ref1}, we define the VSI at the $h$-th bus as
\begin{align}
    &{\rm VSI}_{h} := V_{k_h}^{*4}-4\Big[(P^{L*}_{h}+ a_hP_{h}^{{\rm Ch}*(\tau)})r_{k_hh}-Q^{L*}_{h}X_{k_hh}\Big]^2 \nonumber \\
    & -4\Big[(P^{L*}_{h}+ a_hP_{h}^{{\rm Ch}*(\tau)})r_{k_hh}+Q^{L*}_{h}X_{k_hh}\Big]V_{h}^{(\tau)*2}, \label{vsi_main_exp}
\end{align}
where, $r_{k_hh}$ and $X_{k_hh}:= 2\pi\bar{\omega}l_{k_hh}$ are the resistance and reactance between $k_h^{\text{th}}$ and $h^{\text{th}}$ buses, respectively, and $a_h \in \{0,1\}$ is a binary variable such that $a_h = 1$ if $h \in \mathbb N_{\rm E}$ and $a_h = 0$ otherwise. Note that ${\rm VSI}_{h} \in [0,1]$ from the definition. Since ${\rm VSI}_h$ depends on $i^{*(\tau)}$, $\mu^{(\tau)}$ and $\tau$, the stacked version of ${\rm VSI}_h$ for $h \in \mathbb N_{\rm rec}$ can be written as ${\rm VSI}(i^{*(\tau)}, \mu^{(\tau)},\tau)$. The closer the VSI is to unity, the less susceptible the bus is to a voltage collapse. 
% \red{Therefore, the VSI inherently limits the charging-induced loading and associated voltage-stability degradation, although explicit bus voltage and branch thermal constraints are left for future extensions.}

%  \begin{figure*}[h]
%     \centering
%     \includegraphics[height=10cm, width=18cm]{Problem_framework.eps}
%     \caption{\footnotesize A schematic illustrating how EV owners submit their preferred charging station and time slot, how a central controller evaluates grid conditions and offers incentives to adjust charging rates or reschedule time slots if needed, and how the final, agreed‐upon charging plan is then implemented.}
%     \label{problem_layout_main}
% \end{figure*}

\section{Problem Formulation}\label{sec_prob_formulation}
\subsection{Problem Setup}
\begin{figure}[t]
    \centering
    \includegraphics[width=0.96\linewidth]{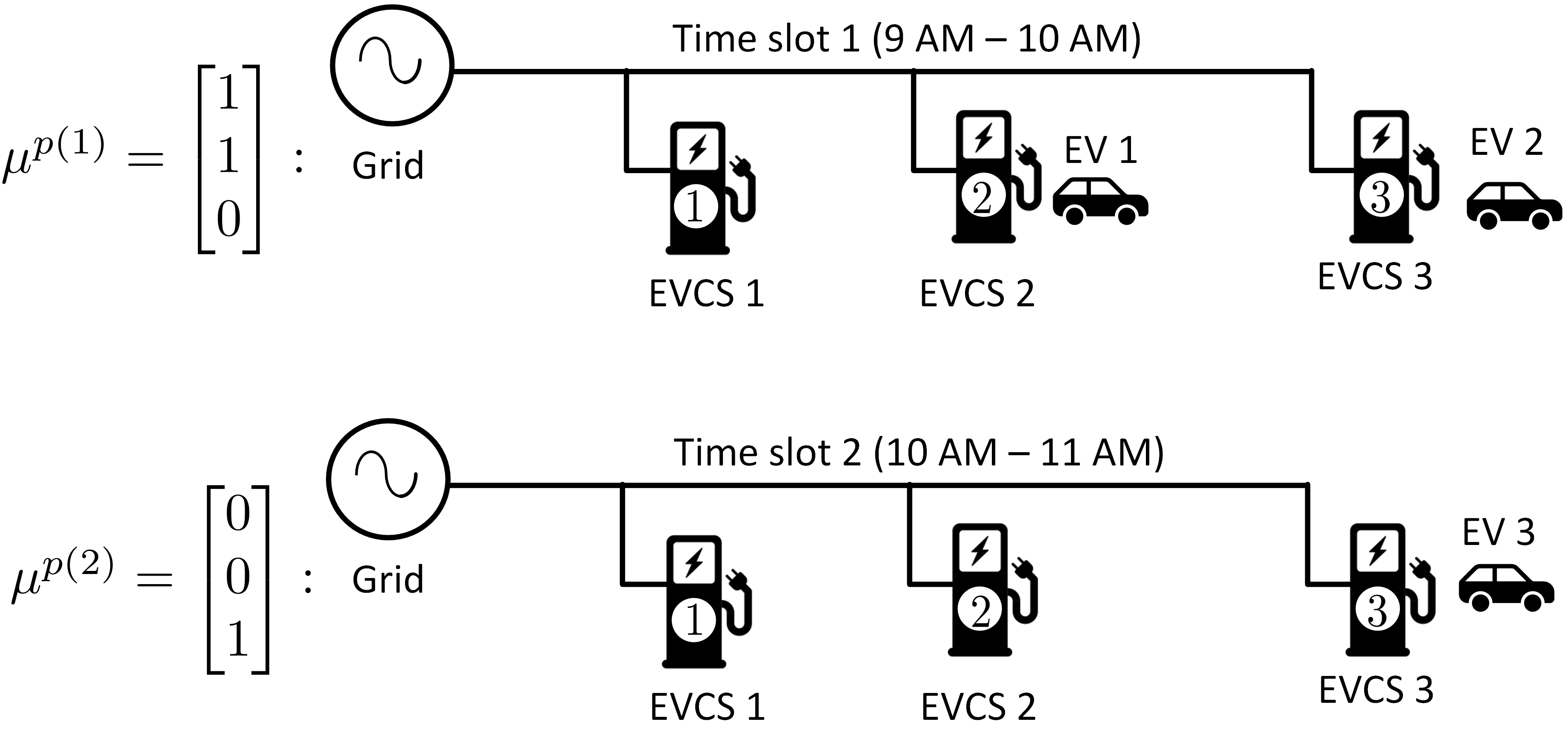}
    \caption{Example of EV assignment to EVCSs based on user preferences $\{e_v^p,\tau_v^p\}_{v\in\mathcal V}$, $\mathcal V=\{1,2,3\}$, $\mathcal E=\{1,2,3\}$. EV 1, 2, and 3 prefer $(e_1^p=2,\tau_1^p=1)$, $(e_2^p=3,\tau_2^p=1)$, and $(e_3^p=3,\tau_3^p=2)$, respectively. The resulting preferred assignment is unique: $\mu^{p(1)}=[1\,1\,0]^\top$ (EVs~1 and~2 at slot~1) and $\mu^{p(2)}=[0\,0\,1]^\top$ (EV~3 at slot~2).}
    \label{mu_assignment}
\end{figure}

We assume that the distribution system operator, referred to here onward as the central point operator (CPO), runs a day-ahead planning problem for determining which EV at any given day and at any given EVCS will charge at which time-slot and for how long. On the day before charging, each EV $v\in\mathcal{V}$ submits the following information: 
\begin{itemize}
    \item[(i)] its preferred EVCS index $e_v^{p}\in\mathcal{E}$, 
    \item[(ii)] its preferred arrival time slot $\tau_v^{p}\in\mathcal{T}$, and 
    \item[(iii)] its energy request $E_v^D$ [kWh]. 
\end{itemize}
The  triple $\{e_v^{p}, \tau_v^{p}, E_v^{D}\}$ for $v\in\mathcal{V}$ constitutes the so-called {\it consumer's objective} \cite{yinling}. It is worth noting that all three quantities dictate the operating point (via power flow) of the grid model, and therefore, the small-signal stability and small-signal performance of  of \eqref{Transfer_func}. Fig.~\ref{mu_assignment} shows a simple example of this triple in which each EV $v$ is connected to $e_v^p$ at time slot $\tau_v^p$.  
In response to this information, the owner of EVCS $e \in \mathcal E$ sends $\{{\sf i}_e^{\max}, {\sf i}_e^{\min}, \pi_e\}$ to the CPO, where ${\sf i}_e^{\max}$ is the maximum charging-current rate defined as ${\sf i}_e^{\max} := P^{\rm max}_{e} / v^{dc*}_{e}$, ${\sf i}_e^{\max}$ is its minimum value, and $\pi_e$ is the per-kWh energy price. Based on the declared $e_v^p$ for each EV $v \in \mathcal V$ and ${\sf i}_e^{\max}$ for $e \in \mathcal E$, the CPO then defines the maximum (resp. minimum) charging current rate {\it per EVs} as $i_v^{\rm max} := {\sf i}_{e_v^p}^{\max}$ (resp. $i_v^{\rm min} := {\sf i}_{e_v^p}^{\min}$). Based on $\{\tau_v^{p}, i_v^{{\rm max}}, i_v^{{\rm min}}\}$ for $v \in \mathcal V$, the CPO determines the optimized charging‑current vector $i^{*(\tau)}$ in \eqref{def_iestar} (the value of whose elements will be less than or equal to the corresponding demanded value $i^{\rm max}$) for each $\tau$), and simultaneously designs a state-feedback controller $u^{(\tau)}(t)$ in \eqref{overall_ss} in a way that their combined effect improves the small-signal dynamic performance of $z^{(\tau)}(t)$. This improvement comes in exchange of monetary incentivization that the CPO must provide to each EV driver at every time slot $\tau$. We present a detailed formulation of the CPO's objective function in the following subsections. 

\vspace{-0.15in}
\subsection{CPO's Objective Function}
%The damping performance can be evaluated as the squared time integral of $z^{(\tau)}(t) := Q\tilde{x}^{(\tau)}(t) + R u^{(\tau)}(t)$ with a state-feedback controller $u^{(\tau)}(t) = -K^{(\tau)}\tilde{x}^{(\tau)}(t)$, driven by the disturbance $x^{(\tau)}(0)-x^{*(\tau)}$, where $\tilde{x}^{(\tau)}(t) := x^{(\tau)}(t)-x^{*(\tau)}$ and $x^{(\tau)}(t)$ follows the nonlinear system $\Sigma^{(\tau)}$ in \eqref{overall_ss}. However, the computation of this metric is intractable due to its high dimensionality and nonlinearity. To enable tractable optimization, we 
We linearize $\Sigma^{(\tau)}$ around the setpoint $\{i^{*(\tau)}, \mu^{(\tau)}\}$ for each time slot $\tau$, and quantify the damping performance of the output variable by the objective function
\begin{align}\label{defJ1}
    J_0(i^{*(\tau)}, \mu^{(\tau)}) := \|G_{i^{*(\tau)},\mu^{(\tau)}}\|_{\mathcal{H}_2}^2, 
\end{align}
where, $G_{i^{*(\tau)},\mu^{(\tau)}}$ is the transfer function from $\hat{x}^{(\tau)}(0)$ to $\hat{z}^{(\tau)}(i^{*(\tau)}, \mu^{(\tau)})$ in \eqref{Transfer_func}, under the state-feedback control
\begin{equation}\label{SFB}
    u^{(\tau)}(t) = -K^{(\tau)}\hat{x}^{(\tau)}(t). 
\end{equation}
%captured as the objective function $J_1$ in \eqref{EV_scheduling_prob}-\eqref{ie_star}.
To keep the VSI at relevant buses to be close to unity following \eqref{vsi_main_exp}, we consider an additional objective function 
\begin{equation}
\hH J_1(i^{*(\tau)},\mu^{(\tau)},\tau) := \|{\bf 1_{|\mathbb{N}_{\rm rec}|}}-{\rm VSI}(i^{*(\tau)}, \mu^{(\tau)},\tau )\|_{\infty,\beta^{\rm VSI}}
\end{equation}
where $\beta^{\rm VSI} \in \mathbb R^{|\mathbb{N}_{\rm rec}|}_{\geq 0}$ is a given constant penalty for the deficit in the VSI. It follows from power flow that both $J_0$ and $J_1$ are monotonically increasing functions of $i_v^{*(\tau)}$, with their minima at $i_v^{*(\tau)}=0$. To avoid this trivial solution, we introduce a third objective function that ensures that $i_v^{*(\tau)}$ does not deviate too much from the demanded current. The optimal (to be designed) and maximum charging power rates for EV $v$ at time slot $\tau$ are $P_{e_v^p}^{*(\tau)} := i_v^{*(\tau)} v^{dc\ast}_{e_v^p}$ and\footnote{In the following we will impose $i_v^{*(\tau)} = 0$ if $\mu^{(\tau)}_v = 0$ as in \eqref{ie_star_1}, thus $P_{e_v^p}^{*(\tau)} = \mu_v^{(\tau)} i_v^{*(\tau)} v^{dc\ast}_{e_v^p}$ holds.} $P_{e_v^p}^{\max} := i_v^{\max} v^{dc\ast}_{e_v^p}$. The charging durations (in minutes) at these rates are $T^{*(\tau)}_{v}:= 60 E_{v}^{D} / P^{*(\tau)}_{e_v^p}$ and $T^{\min}_{v}:= 60 E^{D}_{v} / P^{\max}_{e_v^p}$, respectively. We design an incentive proportional to the time difference $T_v^{*(\tau)}-T_v^{\min}$. To price this extra time, we adopt a ``wait \& save" strategy, similar to ride-sharing protocols (e.g., Lyft; for details, please see \cite{tcst}). Let the charging cost at the demanded rate be $C_v^{D}:=E_v^{D}\,\pi_{e_v^p}$. Under these settings, the incentive payment for wait-and-save can be written as
\begin{equation}\label{incentive_CC}
    \HH M_v^{\rm C}(i_v^{*(\tau)}) := C_v^{D} \frac{T_v^{*(\tau)} - T_v^{\min}}{T_v^{\min}} = \beta_v^{\rm C}\left(\frac{1}{i^{*(\tau)}_v}-\frac{1}{i^{\rm max}_v}\right),
\end{equation}
for $v\in\mathcal V$ with $\beta_v^{\rm C} := i^{\max}_vC_v^D$. Note that $M_v^{\rm C}$ is always non-negative because $i^{*(\tau)}_v \leq i_v^{\rm max}$, which is imposed later. Besides wait-and-save, the CPO may also shift the time-slot for charging any EV from its preferred slot in case the latter is already crowded with too many requests. In other words, the CPO may assign $\tau \not= \tau_v^p$ for EV $v$. This gives rise to another incentive, namely the {\it incentive for time shifting}, which is defined as
\begin{align}\label{incentive_TS}
    M_v^{\rm S}(\tau) := \beta_v^{\rm S} E_v^D |\tau - \tau_v^p|,
\end{align}
for $v\in\mathcal V$, and $\beta_v^{\rm S} \in \mathbb R_{\geq 0}$. The total incentivization objective can be summarized as  
\begin{equation}
    J_{2}(i^{*(\tau)}, \mu^{(\tau)}, \tau) := \sum_{v \in \mathcal V} \mu_{v}^{(\tau)}\left(M_v^{\rm C}(i_v^{*(\tau)})  + M_v^{\rm S}(\tau)\right). \label{defJ2}
\end{equation}
The pre-multiplying factor $\mu_{v}^{(\tau)} \in \{0,1\}$ ensures that the incentivization should be zero when $\mu_{v}^{(\tau)} = 0$, i.e., EV $v$ is not connected to the corresponding EVCS at the time slot $\tau$. By balancing the three objective functions $J_0$, $J_1$ and $J_2$ with constant weights $\gamma_1 \in [0,1]$ and $\gamma_2 \in [0,1]$ such that $\gamma_1 + \gamma_2 \leq 1$, the CPO's objective function can be written as  
\begin{align}\label{CPOobj}
    J^{(\tau)}(&i^{*(\tau)},\mu^{(\tau)}) := (1-\gamma_1-\gamma_2) J_0(i^{*(\tau)}, \mu^{(\tau)}) \nonumber \\
    &+ \gamma_1 J_1(i^{*(\tau)},\mu^{(\tau)},\tau) + \gamma_2 J_{2}(i^{*(\tau)}, \mu^{(\tau)}, \tau).
\end{align}
Changing the weighting coefficients can lead to different Pareto-preferred outcomes, as verified in \cite{tcst}. In our forthcoming simulations, we will only show the case for equal priority of the three objectives for brevity.
% \red{The weighting constants $\gamma_1, \gamma_2$ decide the relative priorities among the system $\mathcal{H}_2$-norm, VSI, and incentivization, and should be selected by the CPO according to the desired tradeoff between dynamic performance, voltage stability, and user compensation. The changing of the weighting coefficients can lead to different Pareto-preferred outcomes, as verified in \cite{tcst}. We only show the equal priority case in the simulation section, and the solutions of other cases are skipped for brevity.}
%where $\gamma_1 \in [0,1]$ and $\gamma_2 \in [0,1]$ such that $\gamma_1 + \gamma_2 \leq 1$ are given constant. 

\noindent \textbf{Constraints:} 
Each EV is assigned to exactly one time slot at a specified EVCS. Thus, the binary variable $\mu_{v}^{(\tau)}$ must satisfy
\begin{align}
    \sum_{\tau\in \mathcal T} \mu_{v}^{(\tau)} = 1, \quad \mu_{v}^{(\tau)} \in \{0,1\}, \quad v \in \mathcal V, \quad \tau \in \mathcal T.\label{one_slot_placement_1}
\end{align}
In addition, each EVCS connects at most one EV during a specified time slot. To represent this, we define the set of all EVs that use EVCS $e$ as $\mathcal V_{e} := \{ v' \in\mathcal V \mid e_{v'}^p = e\}.$ This leads to the following constraint:  
\begin{equation}
\sum_{v\in \mathcal V_e}\mu_v^{(\tau)} \le 1, \quad  \forall e \in \mathcal E, \quad \forall \tau \in \mathcal T. \label{EVCS_cons}
\end{equation}
Moreover, if EV $v$ is not connected to the corresponding EVCS, no charging will occur, whereas if it is connected, then the charging rate must be limited to the allowable range. This constraint can be represented as 
\begin{equation}
    \H \left\{ \hH
    \begin{array}{ll}
    i^{*(\tau)}_{v}= 0, & {\rm if}~ \mu_{v}^{(\tau)} = 0 \\
    i^{*(\tau)}_{v} \in \bigl[i_v^{\min},\,i^{\max}_v\bigr], & {\rm otherwise}
    \end{array}
    \right.\H , ~ v \in \mathcal V, ~\tau \in \mathcal T. \label{ie_star_1}
\end{equation}
%where $i_v^{\rm min} \in \mathbb R_{> 0}$ is constant provided by the corresponding EVCS owner. 

\subsection{Main Problem Statement}
We are now ready to state the main problem. The CPO solves the following day-ahead planning problem: 
\begin{align}
\begin{array}{ll}
     \min_{ \{K^{(\tau)},~i^{*(\tau)},~\mu^{(\tau)}\}_{\tau \in \mathcal T} } & \sum_{\tau \in \mathcal T}J^{(\tau)}(i^{*(\tau)},\mu^{(\tau)}) \\
     {\rm subject~to}& \eqref{Transfer_func}, \eqref{SFB}, \eqref{one_slot_placement_1}, \eqref{EVCS_cons}, \eqref{ie_star_1} 
     \end{array}     
\label{entire_prb}
\end{align}
where, $J^{(\tau)}$ is defined in \eqref{CPOobj}, and $K^{(\tau)}$ is included in \eqref{SFB}.
% After solving for $\mu^{(\tau)}$ from \eqref{entire_prb}, the optimal time slot $\tau^*:=[\tau^{*}_v]^\top_{v\in\mathcal{V}}$ for each EV can be determined as \begin{equation}
% \tau^{*}_v = \sum_{\tau \in \mathcal T} \tau\mu_{v}^{(\tau)}, \quad v \in \mathcal V. 
% \end{equation}
Equation \eqref{entire_prb} is a non-convex optimization problem and involves binary decision variables, so obtaining a global optimum can be difficult. Instead, we numerically find an approximate stationary point of \eqref{entire_prb} using a two-stage greedy framework, which may yield improved values of $\{K^{(\tau)},\, i^{*(\tau)},\, \mu^{(\tau)}\}_{\tau\in\mathcal T}$, as detailed in the next section.
% \red{Equation \eqref{entire_prb} is a non-convex optimization problem and involves binary decision variables, so obtaining a global optimum using the existing methods mentioned in Table \ref{literature_comparison} can be challenging. Instead, we numerically find an approximate stationary point of \eqref{entire_prb} using a two-stage greedy framework, which may yield improved values of $\{K^{(\tau)},\, i^{*(\tau)},\, \mu^{(\tau)}\}_{\tau\in\mathcal T}$, as detailed in the next section.} 

\section{Proposed Method}\label{sec_sol_strategy}
%\subsection{Overview}
Based on the declared customer information $\{e_v^p, \tau_v^{p}\}_{v \in \mathcal V}$, the CPO can uniquely construct the assignment of all EVs for each time slot $\tau$ as
\begin{align} \label{def_mupt}
    \mu_v^{p(\tau)} := \left\{\begin{array}{ll}
    1 & {\rm if}~\tau = \tau_v^{p} \\
    0 & {\rm otherwise}
    \end{array}\right., \quad v \in \mathcal V, \quad \tau \in \mathcal T, 
\end{align}
and $\mu^{p(\tau)} := [\mu_1^{p(\tau)}, \ldots, \mu_{N_{\rm V}}^{p(\tau)}]^{\top} \in \{0,1\}^{N_{\rm V}}$. We refer to $\mu^{p(\tau)}$ as the {\it preferred assignment}, which is assumed to satisfy \eqref{one_slot_placement_1} and \eqref{EVCS_cons}. To solve the optimization problem in \eqref{entire_prb}, we develop the following two-stage algorithm: 
\begin{itemize}
    \item[(1)] Co-optimization under fixed preferred assignments of EVs. That is, we solve \eqref{entire_prb} under $\mu^{(\tau)} \leftarrow \mu^{p(\tau)}$ for each $\tau \in \mathcal T$.  
    \item[(2)] Given the result of Stage (1) as an initial solution, we improve $\{\mu^{(\tau)}, i^{*(\tau)}, K^{(\tau)}\}_{\tau \in \mathcal T}$ using a greedy algorithm. 
\end{itemize}
The purpose of the proposed two-stage approach is therefore computational tractability for day-ahead operation rather than guaranteed global optimality.
% In Stage (1), the EV assignments are fixed at their preferred time slot $\tau_v^p$. This removes the binary assignment search and decomposes the problem into independent slot-wise continuous co-optimization problems. Each subproblem optimizes the charging currents and LQR gain using the gradient-based procedure as shown in \cite{tcst}. On the other hand, Stage (2) performs a local search over time-slot reassignment. For each EV, only currently unoccupied slots at its preferred EVCS are considered. For each candidate reassignment, only the two affected time slots (original and candidate slots) are re-optimized. The candidate producing the largest cost reduction is retained. This substantially reduces the number of full scheduling configurations that must be evaluated compared with exhaustive enumeration of all feasible assignments.
% Stage (1) optimizes the cost under the assumption that each EV is assigned to its declared preferred time slot $\tau_v^p$. Stage (2), on the other hand, seeks additional cost reduction for each EV by reassigning it to another unoccupied slot while re-optimizing $i^{*(\tau)}$ and $K^{(\tau)}$. 
We describe the two stages in detail as follows.

\vspace{-0.11in}
\subsection{Stage (1): Co-optimization under fixed preferred assignment} When $\mu_{v}^{(\tau)} = \mu_{v}^{p(\tau)}$, the EVCS constraint in \eqref{ie_star_1} can be written as 
\begin{equation}
    \left\{ \hH
    \begin{array}{ll}
    i^{*(\tau)}_{v}= 0, & {\rm if}~ \mu_{v}^{p(\tau)} = 0 \\
    i^{*(\tau)}_{v} \in \bigl[i_v^{\rm min}, i^{\max}_v\bigr], & {\rm otherwise}
    \end{array}
    \right.\H, ~v \in \mathcal V, ~\tau \in \mathcal T.  \label{ie_star_1_stg1}
\end{equation}
Although the original problem \eqref{entire_prb} couples all time slots through the assignment constraints \eqref{one_slot_placement_1} and \eqref{EVCS_cons}, in Stage (1), the assignment is fixed. Therefore, the problem can be decomposed into slot-wise independent subproblems. Based on this observation, in Stage (1), the CPO would solve  
\begin{align}\label{stage1_prb}
\begin{array}{ll}
     \min_{K^{(\tau)},~i^{*(\tau)}} & J^{(\tau)}(i^{*(\tau)},\mu^{p(\tau)}) \\
     {\rm subject~to} & \eqref{Transfer_func}, \eqref{SFB}, \eqref{ie_star_1_stg1} \\
     \end{array}
\end{align}
for each $\tau \in \mathcal T$, where $J^{(\tau)}$ is defined in \eqref{CPOobj}. Following our previous work in \cite{tcst}, we use a gradient-based algorithm for solving this optimization, as well be shown in the numerical results in Section V. A brief overview of the method is provided in Remark~\ref{rem1}. 
%\footnote{The reference \cite{tcst} considers a nonlinear system depending on the continuous variable $i^{*}$, and computes the $(A,B)$-pair of an approximate linear system numerically. In our paper, although the system $\Sigma^{(\tau)}$ in \eqref{overall_ss} depends not only on the continuous variable $i^{*(\tau)}$ but also on the binary variable $\mu^{(\tau)}$, note that in Stage~(1) we fix $\mu^{(\tau)}$ by setting $\mu^{(\tau)} \leftarrow \mu^{p(\tau)}$; hence, the method in \cite{tcst} can be applied directly.}. 
Note that the term $\mu_v^{(\tau)}M_v^{\rm S}(\tau)$ in \eqref{defJ2} when $\mu_{v}^{(\tau)} = \mu_{v}^{p(\tau)}$ is always zero, following the definition of $\mu_{v}^{p(\tau)}$ in \eqref{def_mupt}. However, we still include this term because we will use the resultant cost in Stage (2). Let the optimized charging current be denoted as $i^{\star(\tau)}$, and the resultant cost at time slot $\tau$ be denoted as
\begin{align} \label{Jstar_stg1}
   J^{\star(\tau)} := J^{(\tau)}(i^{\star(\tau)},\mu^{p(\tau)}), \quad \tau \in \mathcal T.  
\end{align}
We store 
\begin{align}
    \mathcal{P}^{\star(\tau)} := \Bigl\{i^{\star(\tau)}, \mu^{p(\tau)}, K^{\star(\tau)}, J^{\star(\tau)} \Bigr\}, \quad \tau \in \mathcal T \label{result_stg1}
\end{align}
as the output of Stage (1). %Note that as shown in Remark~\ref{rem1} the state-feedback gain $K^{(\tau)}$ is uniquely determined once $\{i^{*(\tau)},\mu^{p(\tau)}\}$ is fixed. Therefore, we do not store it at this stage.
The entire procedure is summarized in Algorithm \ref{stage_1_alg}. 

\begin{algorithm}[h]\label{stage_1_alg}
\caption{Co-Optimization of Charging Set-Points and Control Gains at the Preferred Time Slots of EVs}
\DontPrintSemicolon
\footnotesize
\KwIn{$\{e_v^p,\tau_v^{p}, E_v^D\}_{v \in \mathcal V}$, $\{{\sf i}_e^{\rm max},{\sf i}_e^{\rm min}, \pi_e\}_{e \in \mathcal E}$,  $Q_{\hat{x}} \geq 0$, $R_{u} > 0$, $\gamma_1,\gamma_2 \in [0,1]$ satisfying $\gamma_1+\gamma_2 \leq 1$, $\beta^{\rm S} \in \mathbb R^{N_{\rm V}}_{\geq 0}$, $\beta^{\rm VSI} \in \mathbb R^{|\mathbb N_{\rm rec}|}_{\geq 0}$. 
Define $i_v^{\rm max} := {\sf i}_{e_v^p}^{\rm max}$, $i_v^{\rm min} := {\sf i}_{e_v^p}^{\rm min}$, $\beta_v^{\rm C}$ in \eqref{incentive_CC}, and $\mu_v^{p(\tau)}$ in \eqref{def_mupt} for each $v$. }

\textbf{Initialize:} For each $\tau \in \mathcal T$ and $v \in \mathcal V$, let $i_v^{*(\tau)} \leftarrow i_v^{\rm max}$ if $\mu_v^{p(\tau)}= 1$, otherwise $i_v^{*(\tau)} \leftarrow 0$. 
%\[
%i_v^{*(\tau)} \leftarrow \left\{
%\begin{array}{ll}
% i_v^{\rm max} & \mu_v^{p(\tau)}= 1, \\
% 0 & {\rm otherwise}.
%\end{array}
%\right. 
%\]

\For{$\tau \in \mathcal T$}
{ 
%Construct $A(i^{*(\tau)},\mu^{p(\tau)})$ and $B(i^{*(\tau)},\mu^{p(\tau)})$ in \eqref{h2pfm_ss_original} by numerical derivative; see \cite{tcst} for the detail. \\
Solve \eqref{stage1_prb} using a gradient-based algorithm in \cite{tcst}; see Remark~\ref{rem1} for a brief overview. \\
Compute $\mathcal P^{\star(\tau)}$ in \eqref{result_stg1}. \\
%For $v$ such that $\tau_v^p=\tau$, assign the optimal tuple for EV $v$:
%        $
%        \mathcal{D}_v \gets \Bigl\{ i_{v,\tau}^{*},\, K^{(\tau)},\, \red{J^{(\tau)}} \Bigr\}.
%        $}
}
\KwOut{$\{ \mathcal{P}^{\star(\tau)} \}_{\tau\in\mathcal T}$ %and $\{\mathcal{D}_v\}_{v=1}^{N_{\rm V}}$ 
} 
\end{algorithm}

\begin{remark}\label{rem1}
    A brief overview for the execution of Step 3 in Algorithm 1 is as follows.  Given $\{i^{*(\tau)}, \mu^{p(\tau)}\}$, we first compute $A(i^{*(\tau)},\mu^{p(\tau)})$ and $B(i^{*(\tau)},\mu^{p(\tau)})$ in \eqref{Transfer_func} by taking the numerical derivative of $\Sigma^{(\tau)}$ in \eqref{overall_ss} along $i^{*(\tau)}$. Next, we design $K^{(\tau)}$ as the LQR for the linearized system $G_{i^{*(\tau)},\mu^{p(\tau)}}$ in \eqref{Transfer_func}. For the closed loop of $G_{i^{*(\tau)},\,\mu^{p(\tau)}}$ and \eqref{SFB}, the exact gradient of $J^{(\tau)}$ with respect to $i^{*(\tau)}$ is not readily available; we, therefore, use an \emph{approximate gradient}, and update $i^{*(\tau)}$ via a line search. By repeating this procedure, we find $i^*(\tau)$. Please see \cite{tcst} for further details onthe implementation of these steps. 
\end{remark}
\begin{remark}\label{rem2}
    Step 1 provides an initial solution for $i^{*(\tau)}$ using the gradient-based method in Step 3. Note that we do not need an initial solution for  $K^{(\tau)}$ because $K^{(\tau)}$ is {\it uniquely}\footnote{The uniqueness of the LQR solution is guaranteed if $(A,\,B)$ is stabilizable and $(A,\,Q_{\hat{x}}^{1/2})$ is detectable. In distribution grid models, these conditions are typically satisfied because the open-loop dynamics near the operating point are asymptotically stable. Both stabilizability and detectability conditions are satisfied for our system model in \eqref{Transfer_func}.} determined from the Algebaric Riccati Equation. 
\end{remark} 

\vspace{-0.11in}

\subsection{Stage (2) - Greedy algorithm for reassignment and re-optimization} 
Once Stage (1) is completed, i.e., $\{i^{\star(\tau)},K^{\star(\tau)}\}_{\tau \in \mathcal T}$ is obtained under the assumption that all EVs are assigned to their preferred time slot, the CPO next sequentially reassigns each EV to the unoccupied time slot that yields the largest reduction in the cost while simultaneously re-optimizing the set $\{i^{\star(\tau)},K^{\star(\tau)}\}$. The algorithm starts with the reassignment of EV 1, followed by EV 2, EV 3, and so on, till all EVs are covered. Let $\mu^{{\rm new}(\tau)} \in \{0,1\}^{N_{\rm V}}$, $i^{{\rm new}{\star}(\tau)}$, $K^{{\rm new}(\tau)}$ and $J^{{\rm new}(\tau)}$ be the reassignment pattern of all EVs, the re-optimized charging current setpoint, the re-optimized state-feedback gain, and the re-optimized cost at time slot $\tau \in \mathcal T$, respectively. We initialize them as
\begin{align}
    &\mu^{{\rm new}(\tau)} := \mu^{p(\tau)}, \quad i^{{\rm new}\star(\tau)} := i^{\star(\tau)}, \nonumber \\
    &K^{{\rm new}(\tau)} := K^{\star(\tau)}, \quad J^{{\rm new}(\tau)} := J^{\star(\tau)}, \label{init_stg2}
\end{align}
for $\tau \in \mathcal T$ by using $\mathcal P^{\star(\tau)}$ in \eqref{result_stg1}. Considering that EVs $\{1, \ldots, v-1\}$ are already reassigned (if $v = 1$ then no EVs are reassigned), we reassign EV $v$ using the steps listed in Algorithm 2. In the following, we refer to step numbers exactly as in Algorithm 2, and describe each step.

\textbf{[Step 4]} For EV $v$, its preferred EVCS is specified as $e_v^p\in\mathcal E$ as described in Section \ref{sec_prob_formulation}. However, note that other EVs may use the same EVCS in different time slots. The set of all EVs that use $e_v^p$ is denoted as $\mathcal V_{e_v^p}$. We define the set of feasible (unoccupied) slots at its preferred EVCS as 
\begin{equation}\label{def_feasset}
\H \mathcal T_v^{\rm feas} := \Bigl\{ \tau\in\mathcal T \;\big|\; \tau\neq \tau_v^p ~\land \;
\mu_{v'}^{{\rm new}(\tau)}=0, \forall v' \in \mathcal V_{e_v^p} \Bigr\},
\end{equation}
where the operation ``$\mu_{v'}^{{\rm new}(\tau)}=0, \forall v' \in \mathcal V_{e_v^p}$" enumerates all currently unoccupied slots at the target EVCS $e_v^p$ under the latest reassignment pattern. 

\textbf{[Step 6]} Let $\tau^{c}_v \in \mathcal T_v^{\rm feas}$, which is a candidate time slot for EV $v$. Suppose the time slot of EV $v$ is hypothetically changed from its preferred time slot $\tau_v^{p}$ to the candidate slot $\tau^{c}_v$. Then the latest reassignment vector $\mu^{{\rm new}(\tau)}$ for $\tau \in \{\tau_v^p, \tau^c_v\}$ is updated as 
\begin{align}
\mu^{c(\tau_v^c)} = \mu^{{\rm new}(\tau_v^c)} + \mathbf e_{v}, \quad \mu^{c(\tau_v^p)} = \mu^{{\rm new}(\tau_v^p)} - \mathbf e_{v}, \label{def_newmu}
\end{align}
where $\mathbf e_v \in \mathbb R^{N_{\rm V}}$ denotes the $v$-th column vector of $I_v$. Note that $\mu^{{\rm new}(\tau)}$ for other time slots is not changed.

\textbf{[Step 7]} For $\tau \in \{\tau_v^p, \tau^c_v\}$, since the reassignment is different from $\mu^{{\rm new}(\tau)}$, the optimization result will be changed. Therefore, we re-solve 
\begin{align}\label{stage2_prb}
\begin{array}{ll}
     {\rm min}_{K^{(\tau)},~i^{*(\tau)}} & J^{(\tau)}(i^{*(\tau)},\mu^{c(\tau)}) \\
     {\rm subject~to} & \eqref{Transfer_func}, \eqref{SFB}, \eqref{ie_star_1_stg1} \\
     \end{array}, \quad \tau \in \{\tau_v^{p}, \tau_v^{c}\},
\end{align}
by using a gradient-based algorithm as in \cite{tcst}, where the initial solution of $i^{*(\tau)}$ is now chosen as $i^{{\rm new}*(\tau)}$. 

\textbf{[Step 8]} For $\tau \in \{\tau_v^{p}, \tau_v^{c}\}$, let the optimized variables be denoted as $\{\tilde{i}^{\star(\tau)},\tilde{K}^{\star(\tau)}\}$, and the new cost be denoted as 
\begin{align} \label{Jstar_stg2}
   \tilde{J}^{\star(\tau)} := J^{(\tau)}(\tilde{i}^{\star(\tau)},\mu^{c(\tau)}). 
\end{align}
Note here that the term $\mu_v^{(\tau)}M_v^{\rm S}(\tau)$ in \eqref{defJ2} when $\mu_{v}^{(\tau)} = \mu_{v}^{c(\tau)}$ and $\tau = \tau_v^{c}$ is {\it not} zero. In other words, $\tilde{J}^{\star(\tau_v^c)}$ includes the incentivization cost caused by the time shifting.  

\textbf{[Step 9]} The total cost variation caused by the reassignment can be defined as the sum of cost variations at each time slot $\tau_v^{p}$ and $\tau^c_v$, i.e.,  
\begin{equation}\label{def_deltatau}
    \H \Delta(\tau_v^{c}) := \left(\tilde{J}^{\star(\tau_v^p)} - J^{{\rm new}(\tau_v^p)}\right) + \left(\tilde{J}^{\star(\tau_v^c)} - J^{{\rm new}(\tau_v^c)}\right),
\end{equation}
where the first term represents the cost variation caused by disconnecting EV $v$ from EVCS $e_v^p$ at its preferred time slot $\tau_v^p$ under the latest reassignment, while the second term represents the cost variation caused by connecting it to the same EVCS at the new time slot $\tau_v^c$ under the latest reassignment. 

\textbf{[Steps 10-12]} Repeating the above procedure for each $\tau_v^{c}\in\mathcal T_v^{\rm feas}$, we determine the time slot $\tau_v^{\rm new}$ satisfying 
\begin{align} 
    \Delta(\tau_v^{\rm new}) \leq \Delta(\tau_v^c),~\forall \tau_v^{c}\in\mathcal T_v^{\rm feas}, \label{tau_new1} 
\end{align}
if there exists $\tau_v^c$ such that $\Delta(\tau_v^c) < 0$ \footnote{If there exist multiple $\tau_v^{\rm new}$ satisfying \eqref{tau_new1}, we choose the one randomly among those. }. In addition, we update the decision variables and cost as 
\begin{align}
    &\mu^{{\rm new}(\tau)} \leftarrow \mu^{c(\tau)}, \quad i^{{\rm new}\star(\tau)} \leftarrow \tilde{i}^{\star(\tau)}, \label{tau_new2} \\
    &K^{{\rm new}(\tau)} \leftarrow \tilde{K}^{\star(\tau)}, \quad J^{{\rm new}(\tau)} \leftarrow \tilde{J}^{\star(\tau)}, \label{tau_new3}
\end{align}
for $\tau \in \{\tau_v^{p}, \tau_{v}^{\rm new}\}$. If not, (i.e., if all candidates increase the cost), then we retain $\tau_v^{p}$, $\mu^{{\rm new}(\tau_v^p)}$, $i^{{\rm new}\star(\tau_v^p)}$, $K^{{\rm new}(\tau_v^p)}$, and $J^{{\rm new}(\tau_v^p)}$. 

We repeat these steps for all $v\in\mathcal V$. The entire procedure is summarized in Algorithm~2, where $\tau_v^{\star}$ denotes the reassigned time slot of EV $v$.

\begin{algorithm}[h]\label{stage_2_alg}
\caption{Greedy Algorithm for Reassignment and Re-optimization}
\DontPrintSemicolon
\footnotesize
\KwIn{Same inputs in Algorithm \ref{stage_1_alg}. Output $\{ \mathcal{P}^{\star(\tau)} \}_{\tau\in\mathcal T}$  by Algorithm \ref{stage_1_alg}. 
}
\textbf{Initialize:} $\{\mu^{{\rm new}(\tau)}, i^{{\rm new}\star(\tau)}, K^{{\rm new}(\tau)}, J^{{\rm new}(\tau)}\}_{\tau \in \mathcal T}$ by \eqref{init_stg2}. In addition, perform a random permutation of $\{1,2,\ldots,N_{\rm V}\}$ and assign the resulting indices as the EV identification numbers. \\
\For{$v \in \mathcal V$}
{ 
Construct the set of feasible slots $\mathcal T_v^{\rm feas}$ by \eqref{def_feasset}. \\
\For{$\tau_v^{c} \in \mathcal T_v^{\rm feas}$} 
{ 
Construct $\mu^{c(\tau_v^c)}$ and $\mu^{c(\tau_v^p)}$ by \eqref{def_newmu}. \\
Solve \eqref{stage2_prb} for $\tau \in \{\tau_v^{p}, \tau_v^{c}\}$. \\
Compute $\tilde{J}^{\star (\tau)}$ in \eqref{Jstar_stg2} for $\tau \in \{\tau_v^{p}, \tau_v^{c}\}$. \\
Compute $\Delta(\tau_v^c)$ in \eqref{def_deltatau}. \\
}

\eIf{ there exists $\tau_v^c$ such that $\Delta(\tau_v^c) < 0$}{ Determine $\tau_v^{\rm new}$ satisfying \eqref{tau_new1}  and update $\{\mu^{{\rm new}(\tau)}, i^{{\rm new}\star(\tau)}, K^{{\rm new}(\tau)}, J^{{\rm new}(\tau)}\}_{\tau \in \{\tau_v^{p}, \tau_{v}^{\rm new}\}}$ by \eqref{tau_new2}-\eqref{tau_new3}.\\

}{Keep $\tau_v^{p}$, $\mu^{{\rm new}(\tau_v^p)}$, $i^{{\rm new}\star(\tau_v^p)}$, $K^{{\rm new}(\tau_v^p)}$, and $J^{{\rm new}(\tau_v^p)}$.\\
}
}

\KwOut{Overwrite $\mu^{\star(\tau)} \leftarrow \mu^{{\rm new}(\tau)}$, $i^{\star(\tau)} \leftarrow i^{{\rm new}(\tau)}$, $K^{\star(\tau)} \leftarrow K^{{\rm new}(\tau)}$, $J^{\star(\tau)} \leftarrow J^{{\rm new}(\tau)}$. Let $\tau_v^{\star} := \sum_{\tau \in \mathcal T}\mu_v^{\star(\tau)}\tau$. Output $\{\mu^{{\star}(\tau)}, i^{\star(\tau)}, K^{\star(\tau)}, J^{\star(\tau)}\}_{\tau \in \mathcal T}$ and $\{\tau_v^{\star}\}_{v\in \mathcal V}. $
} 
\end{algorithm}

\begin{remark}\label{rem_random}
In Step 1 of Algorithm 2, we randomly assign ID numbers to the EVs to ensure fairness of the reassignment. %Because the algorithm is sequential, the reassignment could otherwise be biased across EVs. To prevent this, we randomize the ordering before performing the reassignment. 
\end{remark}

%\begin{remark}
%    Extension to $\tau_v^{p}$ is not scalar. 
%\end{remark}

\subsection{Practical Implementation of Algorithms 1-2}\label{sec_iv}
As mentioned in Section III, EV drivers and EVCS owners submit their respective information, namely $\{e_v^{p}, \tau_v^{p}, E_v^{D}\}_{v \in \mathcal V}$ and $\{{\sf i}_e^{\max}, {\sf i}_e^{\min}, \pi_e\}_{e \in \mathcal E}$ one day ahead of time. Based on this information, the CPO runs  Algorithm~\ref{stage_1_alg} to compute slot-wise optimized charging currents at the preferred assignment $\tau_v^p$, and subsequently runs Algorithm~\ref{stage_2_alg} to obtain the optimized charging rates $i^{\star(\tau)}$, state-feedback gain $K^{\star(\tau)}$, reassignment $\mu^{\star(\tau)}$, and the reassigned time slots $\{\tau_v^{\star}\}_{v\in \mathcal V}$. Using these results, the CPO computes the incentives as follows. 

\begin{figure}[t]
    \centering
    \includegraphics[width=0.45\textwidth]{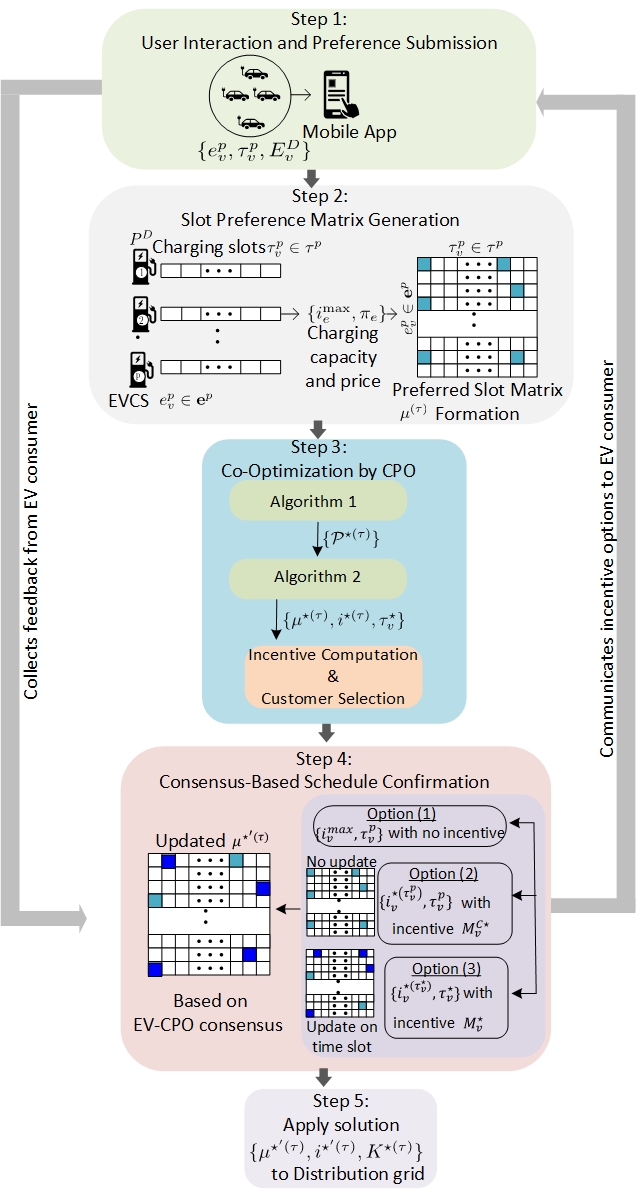}
    % \caption{A schematic illustration of the proposed charging control, scheduling, and incentivization.}
    \caption{A schematic illustrating how EV owners submit their preferred charging station and time slot, how a proposed strategy evaluates grid conditions and offers incentives to adjust charging rates or reschedule time slots if needed, and how the final, agreed‐upon charging plan is implemented into distribution grid.}
    \label{problem_layout_main}
\end{figure}

\textbf{Incentive Computation:} For EV $v$, the optimized charging current is $i_v^{\star(\tau_v^{\star})}$. The incentive for wait-and-save and time-shifting are determined, respectively, as 
\begin{align}
   M_v^{\rm C\star} := M_v^{\rm C}(i_v^{\star(\tau_v^{\star})}), \quad M_v^{\rm S\star} := M_v^{\rm S}(\tau_v^{\star}), \quad v \in \mathcal V, \label{in1}
\end{align}
where $M_v^{\rm C}(\cdot)$ and $M_v^{\rm S}(\cdot)$ are defined in \eqref{incentive_CC}-\eqref{incentive_TS}. Subtracting these incentives from the original charging cost $C_v^D$, the \emph{net} amount of money offered to EV $v$ is then calculated as 
\begin{align}
    C^{\rm net}_{v} := C^{D}_{v}-M^{\star}_v, \quad  M^{\star}_v := M_v^{\rm C\star} + M_v^{\rm S\star}. \label{in2}
\end{align}
The calculations in \eqref{in1} and \eqref{in2} can be integrated into a mobile app, enabling direct communication between the CPO and the EV drivers. The owner of every EV $v$ can view three offers from the CPO, and choose one of the following:
\begin{enumerate}
    \item[(1)] Charge at the consumer’s preferred rate $i_v^{\max}$ at the preferred time slot $\tau_v^p$ with no incentive;
    \item[(2)] Charge at the optimized rate $i_v^{\star(\tau_v^p)}$ at the preferred time slot with the incentive $M_v^{\mathrm{C}\star}$;
    \item[(3)] Charge at the optimized rate $i_v^{\star(\tau_v^\star)}$ at the reassigned time slot with the total incentive $M_v^{\star}$.
\end{enumerate}
Based on each selection, the CPO must update the reassignment $\mu^{\star(\tau)}$ and the charging current $i^{\star(\tau)}$ for $\tau \in \mathcal T$. The updated quantities are denoted as $\{\mu^{\star'(\tau)},\, i^{\star'(\tau)}\}_{\tau \in \mathcal T}$. Finally, for the time slot $\tau$, the CPO implements the charging time assignment, wait-and-save strategy, and state feedback control using the triple $\{\mu^{\star'(\tau)},\, i^{\star'(\tau)},\, K^{\star(\tau)}\}_{\tau \in \mathcal T}$. Figure~\ref {problem_layout_main} illustrates EV-CPO interaction and the implementation of the proposed method.
In summary, the current incentive formulation can be viewed as a stability-aware correction layer that complements, rather than replaces, economically driven EV scheduling, which includes charging station revenue, network operation cost, and so on.

\section {Numerical Results}\label{sec_results}
\begin{figure}[t]
    \centering
    \includegraphics[width=0.48\textwidth]{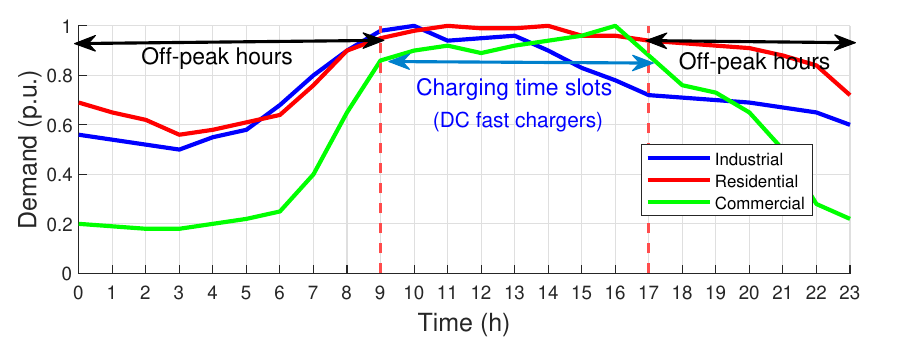}
    \caption{Typical daily load profiles for industrial (blue), residential (red), and commercial (green) load models. }
    \label{load_profile}
\end{figure}

We validate our proposed method using the IEEE 33-bus radial feeder model with 10 EVCSs. Details of this model can be found in \cite{33_bus}.
% The testbed operates at 12.66 kV, has 33 buses, and supports a 3.715 MW non-EV load, which consists of industrial, commercial, and residential loads, collectively referred to as non-EV loads. Please see \cite{33_bus} for the detailed information, such as $\mathbb N_{\rm L}$.
The EVCS buses are represented as $\mathbb{N}_{\rm E}:=\{3,5,9,11,15,17,19,21,26,32\}$, while the EVCS themselves are indexed as $\mathcal{E}=\{1,2,\ldots,10\}$. For instance, EVCS at bus 3 is indexed as 1, that at bus 5 is numbered as 2, and so on. The model parameters of each EVCS are considered following \cite{tcst}. For EVCS $e \in \mathcal E$, its maximum capacity $P_e^{\rm max}$ is given as $\{50,75,150,175,100, 75,50,100,150,50\}$ kW, or equivalently as ${\sf i}_e^{\max} =\{62.5,93.75,187.5,218.75,125,93.75,62.5,$ $125,187.5,62.5\}$ Amps. Let ${\sf i}_e^{\min} =0.85{\sf i}_e^{\max}$ Amps, and its charging price be $\pi_e=\$0.45$/kWh if $P^{\rm max}_e =50$ kW and \$0.5/kWh otherwise. Following \cite{tcst}, let $\beta_v^{\rm S}=\$0.04/\text{kWh}$ in \eqref{incentive_TS} if $P_e^{\rm max}\in \{50,75\}$ kW and $\beta_v^{\rm S}=\$0.02/\text{kWh}$ otherwise. We consider $\gamma_1=\gamma_2=0.33$, $Q_{\hat{x}} = 0.0001 I$ in \eqref{defxhat}, $R_u= I_{10} \otimes {\rm diag}(100, 100, 1)$, $\beta^{\rm VSI}=0.005 {\bf 1}_{|\mathbb N_{\rm rec}|}$, and $v_e^{dc*}=800V$ for any $e$. All simulations are done using MATLAB/Simulink\textsuperscript{\textregistered}  on a PC equipped with an Intel(R) Core(TM) i7-11390H @ 3.40GHz processor and 16GB  RAM. Please note that although the two-stage algorithm is solved using the small-signal model, all simulations reported in the following subsections pertain to the nonlinear model of the 33-bus test system.
Fig.~\ref{load_profile} shows the load profile for three typical non-EV loads, namely - industrial, commercial, and residential. The loads peak between hours 09:00 and 17:00, so we partition this window into eight one-hour slots. Therefore, 09:00-10:00 hours is indexed as time slot 1, while 16:00-17:00 hours is indexed as time slot 8, with $\mathcal T = \{1,\dots,8\}$. 
\begin{figure}[t]
    \centering
    \includegraphics[width=0.48\textwidth]{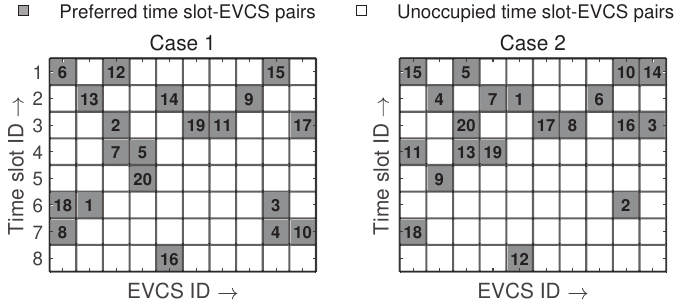}
    \caption{EV consumers' preferred charging slots at specified EVCSs under two cases: (1) random-hour charging, and (2) early-morning charging. }
    \label{choice}
\end{figure}
% We assume that $20$ EVs will be connected to the grid for charging over the day spanning 9:00 hours to 17:00 hours. i.e., $\mathcal{V}=\{1,2,\ldots,20\}$. The ID numbers of EVs are randomly assigned per planning period, as mentioned in Remark \ref{rem_random}. For $v \in \mathcal V$, let $E_v^D = 45\,\text{kWh}$ if $P^{\text{max}}_{e_v^p} \in \{50, 75\}\,\text{kW}$ and $E_v^D = 90\,\text{kWh}$ otherwise. 
% \red{As shown  in Fig.~\ref{problem_layout_main}, every consumer sends its preferred information $\{e_v^p, \tau_v^p, E_v^D\}$ to the CPO one day before. To account for the uncertainty of charging-preference patterns, we show two scenarios in Fig.~\ref{choice}, such as random-hour and early-morning charging scenarios, and demonstrate the behavior of the proposed framework under different user-preference distributions.
% % In Fig.~\ref{choice}, we show two cases for $\{e_v^p, \tau_v^p\}_{v \in \mathcal V}$. The first case shows the cars charging at random hours, while the second case shows early-morning charging preference by the drivers. 
% In the figure, each gray box indicates the preferred charging slot $\tau_v^p \in \mathcal{T}$ at a particular EVCS $e_v^p \in \mathcal{E}$ chosen by EV $v \in \mathcal{V}$. 
% For example, the top-left gray box in Case~1 indicates that EV~6 prefers EVCS~1 at time slot~1. 
% Numerical results of our proposed algorithms are presented as follows.}
We assume that $20$ EVs will be connected to the grid for charging over the day spanning 9:00 hours to 17:00 hours. i.e., $\mathcal{V}=\{1,2,\ldots,20\}$. The ID numbers of EVs are randomly assigned per planning period, as mentioned in Remark \ref{rem_random}. For $v \in \mathcal V$, let $E_v^D = 45\,\text{kWh}$ if $P^{\text{max}}_{e_v^p} \in \{50, 75\}\,\text{kW}$ and $E_v^D = 90\,\text{kWh}$ otherwise. 
Every consumer sends its preferred information $\{e_v^p, \tau_v^p, E_v^D\}$ to the CPO one day before. To account for the uncertainty of charging preference, we consider two cases as in Fig.~\ref{choice}, namely, random-hour charging and early-morning charging. %, and demonstrate the behavior of the proposed framework under different user-preference distributions.
% In Fig.~\ref{choice}, we show two cases for $\{e_v^p, \tau_v^p\}_{v \in \mathcal V}$. The first case shows the cars charging at random hours, while the second case shows early-morning charging preference by the drivers. 
% Numerical results of our proposed algorithms are presented as follows.

\subsection{Co-optimization under fixed preferred assignment}
We apply Algorithm 1 to compute the optimized charging current setpoints ${i}^{\star(\tau)}$, the control gain $K^{\star(\tau)}$, and the objective function $J^{\star(\tau)}$ for each of the eight time slots, as shown in Fig.~\ref{choice}. For Case 1 (random hour charging), in time slot~1, the original charging currents of EVCS 1, 3, and 9 are $62.5$ A, $187.5$ A, and $187.5$ A, while the optimized values are $59.51$ A, $181.38$ A, and $182.43$ A, respectively. For Case 2 (early morning charging), in time slot~1, the original charging currents of EVCS 1, 3, 9, and 10 are $62.5$ A, $187.5$ A, $187.5$ A, and $62.5$ A, while the optimized values are $58.31$ A, $182.11$ A, $181.4$ A, and $57.92$ A, respectively. Algorithm \ref{stage_1_alg} optimizes $\{{i}^{\star(\tau)},K^{\star(\tau)}, J^{\star(\tau)}\}$ for all $20$ EVs at every time slot for both cases. The resultant information is then utilized and re-optimized as follows.

\subsection{Greedy Algorithm for Reassignment and Re-optimization}
\begin{figure}[h]
    \centering
    \includegraphics[width=0.47\textwidth]{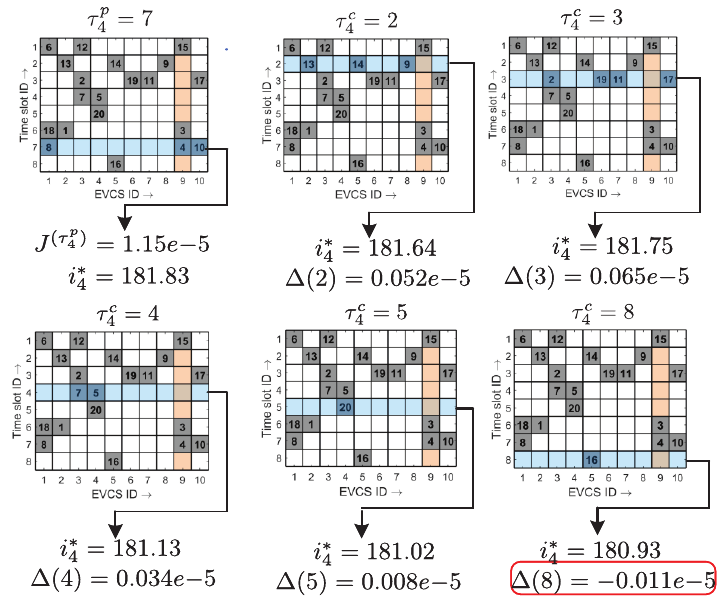}
    \caption{ Illustration of the step-by-step time-slot search for EV 4 in Case 1. In each subfigure, gray/white boxes indicate occupied/unoccupied slot pairs; the blue box denotes the target time slot, and the orange box marks the slot that is currently being tested for shifting.}
    \label{timeslot variation}
\end{figure}

\begin{figure}[t]
    \centering
    \includegraphics[width=0.49\textwidth]{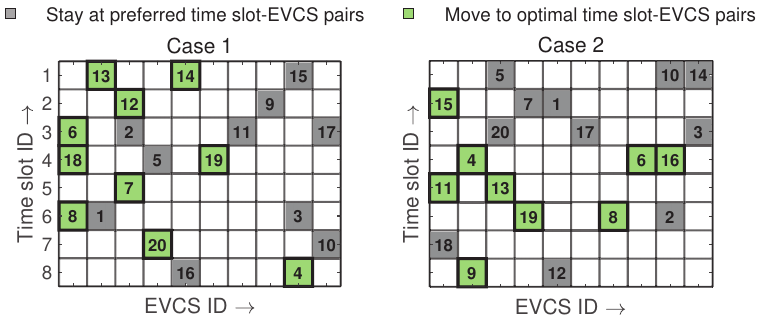}
    \caption{Day-ahead EV charging preferences (gray) and suggested optimized time slots (green) for two cases. In total, only 10 and 9 of the 20 EVs were shifted from their preferred time slots in Case~1, and 2, respectively. }
    \label{optimal_choice}
\end{figure}

In this stage, we reassign $\mu^{\star(\tau)}$ and re-optimize $\{i^{\star(\tau)},\,K^{\star(\tau)}\}$. For Case 1, we start by testing the reassignment of EV~1, and find that the preferred time slots for EVs 1, 2, and 3 are also the optimized time slots. Fig.~\ref{timeslot variation} shows the greedy slot-search process (Steps~3-12 of Algorithm~2) for EV 4. The top-left assignment map shows the initial assignment. We hypothetically move EV~4 from its preferred slot $\tau_4^c=7$ to $\tau_4^c=2$ and compute the resulting cost difference. In slot~7, originally EVs $\{4,\,8,\,10\}$ are connected at their preferred EVCSs (as shaped by the blue box), after the time shift, two EVs $\{8,\,10\}$ are connected, thereby changing the cost at slot~7. Similarly, in slot~2, originally three EVs $\{9,13,14\}$ are connected; after moving EV~4 into slot~2, four EVs $\{4,9,13,14\}$ are connected, thereby changing the cost at slot 2. The sum of these changes yields the cost difference $\Delta(2)=0.052 \times 10^{-5}$ (as defined in \eqref{def_deltatau}). We repeat the same computation for all feasible time slots $\tau^c_4=\{3,4,5,8\}$ because slots $6$ and $7$ are occupied, and move EV~4 to the time slot that minimizes $\Delta(\tau_v^c)$ (slot~8 in this example). We then update the assignment map with EV~4 moved to slot~8 and apply the same procedure to subsequent EVs. %Note that at the end of each reassignment, we update $i^{\star(\tau)}$ and $K^{\star(\tau)}$ to the latest values obtained from recomputing the cost in the affected time slots. 
The results for all $20$ EVs are shown in Fig.~\ref{optimal_choice}. 
% The computation time for Case 1 is 195 seconds, as is similar for Case~2, which is compatible with day-ahead planning.
% \red{Amit - Since we assume that the EV IDs have already been randomly assigned, the reassignment should be carried out for EVs $1,2,\ldots,20$ in that order. Accordingly, the ID numbers in all figures should be updated.} -\blue{I am quite unclear about this, need to discuss with you. In Figure 5, I show the process of EV 6 of case 1. Do you want me to show the process of EV 1 (which sits at EVCS 2 at time slot 6)? }

%The results for all $20$ EVs of Case~1 are shown in Fig.~\ref{optimal_choice}, which takes approximately 195 seconds to compute, as is similar for Case~2. The proposed optimization recommends shifting certain EVs to alternative slots at the same EVCS, which are marked as lime boxes. Crucially, the algorithm is aware of each EVCS’s position in the considered IEEE 33 bus testbed, since VSI degrades more when heavy loads are placed on remote buses than on those near the substation (bus 1). 
%Consequently, the algorithm preferentially moves EVs away from EVCSs at the network edges (i.e., mainly the EVCS at bus 9-17). Therefore, most time slot shifts occur at EVCSs 2-6, where the impact of charging load is more significant on VSI than the EVCSs near the substation (bus 1).
% In contrast, the EV loads at EVCS closer to bus 1 remain largely unchanged, as their local VSI is robust even under peak charging conditions.
%For example, in Case~2, EV 16 at EVCS 9 is reassigned from its preferred time slot 3 to 4 to improve VSI across all eight time slots. In total, for Case 1, only 10 out of the 20 EVs, and for Case 2, only 9 out of the 20 EVs were shifted from their preferred time slots. %Thus, the algorithm maximizes VSI and damping performance while ensuring that customer preferences are met as much as possible. 

\begin{table*}[t]
\scriptsize
\centering
\caption{Comparison of three different options via leveraging the proposed approach for case 1. Here, $g_1^{(\tau)} := \|G_{i^{\rm max}_v,\mu^{p(\tau)}}\|_{\mathcal H_2}$, $g_2^{(\tau)} := \|G_{i^{\star'(\tau)},\mu^{p(\tau)}}\|_{\mathcal H_2}$, and $g_3^{(\tau)} := \|G_{i^{\star'(\tau)},\mu^{\star'(\tau)}}\|_{\mathcal H_2}$, where $G$ is defined in \eqref{Transfer_func}, and ${\underbar{$\mathrm V$}}:={\rm VSI}_{\min}$. } 
\label{sumtab4}
\begin{tabular}{|c|ccccc|cccccc|ccccccc|}
\hline
\multirow{2}{*}{$\tau$} & \multicolumn{5}{c|}{\begin{tabular}[c]{@{}c@{}}At the preferred charging rate and time slot\\ (Option 1)\end{tabular}}                                                                    & \multicolumn{6}{c|}{\begin{tabular}[c]{@{}c@{}}At the preferred time slot with optimized charging rate\\ (Option 2)\end{tabular}}                                                                                                             & \multicolumn{7}{c|}{\begin{tabular}[c]{@{}c@{}}At optimized charging rate and time slot\\ (Option 3)\end{tabular}}                                                                                                                                                                \\ \cline{2-19} 
                        & \multicolumn{1}{c|}{$v$} & \multicolumn{1}{c|}{${ i}^{\rm max}_v$} & \multicolumn{1}{c|}{$g_1^{(\tau)}$}           & \multicolumn{1}{c|}{${\underbar{$\mathrm V$}}$}              & $\hspace{-5pt}M^{\rm C\star}$    & \multicolumn{1}{c|}{$v$} & \multicolumn{1}{c|}{$ i^{\star^\prime(\tau)}_v$} & \multicolumn{1}{c|}{$g_2^{(\tau)}$}           & \multicolumn{1}{c|}{${\underbar{$\mathrm V$}}$}              & \multicolumn{1}{c|}{$M_v^{\rm C\star}$} & $\hspace{-5pt}M^{\rm C\star}$        & \multicolumn{1}{l|}{$\tau$}             & \multicolumn{1}{c|}{$v$} & \multicolumn{1}{c|}{$i^{\star^{\prime}(\tau)}_v$} & \multicolumn{1}{c|}{$g_3^{(\tau)}$}   & \multicolumn{1}{c|}{${\underbar{$\mathrm V$}}$}              & \multicolumn{1}{c|}{$M_v^\star$} & $\hspace{-5pt}M^\star$               \\ \hline
\multirow{3}{*}{1}      & \multicolumn{1}{c|}{6}   & \multicolumn{1}{c|}{62.5}                  & \multicolumn{1}{c|}{\multirow{3}{*}{1.09e-3}} & \multicolumn{1}{c|}{\multirow{3}{*}{0.681}} & \multirow{20}{*}{\hspace{-5pt}0} & \multicolumn{1}{c|}{6}   & \multicolumn{1}{c|}{59.51}                     & \multicolumn{1}{c|}{\multirow{3}{*}{5.98e-4}} & \multicolumn{1}{c|}{\multirow{3}{*}{0.709}} & \multicolumn{1}{c|}{1.02}               & \multirow{20}{*}{\hspace{-5pt}31.51} & \multicolumn{1}{c|}{\multirow{3}{*}{1}} & \multicolumn{1}{c|}{13}  & \multicolumn{1}{c|}{89.37}                      & \multicolumn{1}{c|}{\multirow{3}{*}{4.35e-4}} & \multicolumn{1}{c|}{\multirow{3}{*}{0.718}} & \multicolumn{1}{c|}{2.90}        & \multirow{20}{*}{\hspace{-5pt}50.96} \\ \cline{2-3} \cline{7-8} \cline{11-11} \cline{14-15} \cline{18-18}
                        & \multicolumn{1}{c|}{12}  & \multicolumn{1}{c|}{187.5}                 & \multicolumn{1}{c|}{}                         & \multicolumn{1}{c|}{}                       &                     & \multicolumn{1}{c|}{12}  & \multicolumn{1}{c|}{181.38}                    & \multicolumn{1}{c|}{}                         & \multicolumn{1}{c|}{}                       & \multicolumn{1}{c|}{1.52}               &                         & \multicolumn{1}{c|}{}                   & \multicolumn{1}{c|}{14}  & \multicolumn{1}{c|}{120.54}                     & \multicolumn{1}{c|}{}                         & \multicolumn{1}{c|}{}                       & \multicolumn{1}{c|}{3.46}        &                         \\ \cline{2-3} \cline{7-8} \cline{11-11} \cline{14-15} \cline{18-18}
                        & \multicolumn{1}{c|}{15}  & \multicolumn{1}{c|}{187.5}                 & \multicolumn{1}{c|}{}                         & \multicolumn{1}{c|}{}                       &                     & \multicolumn{1}{c|}{15}  & \multicolumn{1}{c|}{182.43}                    & \multicolumn{1}{c|}{}                         & \multicolumn{1}{c|}{}                       & \multicolumn{1}{c|}{1.25}               &                         & \multicolumn{1}{c|}{}                   & \multicolumn{1}{c|}{15}  & \multicolumn{1}{c|}{182.23}                     & \multicolumn{1}{c|}{}                         & \multicolumn{1}{c|}{}                       & \multicolumn{1}{c|}{1.25}        &                         \\ \cline{1-5} \cline{7-11} \cline{13-18}
\multirow{3}{*}{2}      & \multicolumn{1}{c|}{13}  & \multicolumn{1}{c|}{93.75}                 & \multicolumn{1}{c|}{\multirow{3}{*}{1.14e-3}} & \multicolumn{1}{c|}{\multirow{3}{*}{0.701}} &                     & \multicolumn{1}{c|}{13}  & \multicolumn{1}{c|}{89.38}                     & \multicolumn{1}{c|}{\multirow{3}{*}{5.87e-4}} & \multicolumn{1}{c|}{\multirow{3}{*}{0.713}} & \multicolumn{1}{c|}{1.71}               &                         & \multicolumn{1}{c|}{\multirow{2}{*}{2}} & \multicolumn{1}{c|}{12}  & \multicolumn{1}{c|}{180.92}                     & \multicolumn{1}{c|}{\multirow{2}{*}{3.58e-4}} & \multicolumn{1}{c|}{\multirow{2}{*}{0.724}} & \multicolumn{1}{c|}{3.46}        &                         \\ \cline{2-3} \cline{7-8} \cline{11-11} \cline{14-15} \cline{18-18}
                        & \multicolumn{1}{c|}{14}  & \multicolumn{1}{c|}{125}                   & \multicolumn{1}{c|}{}                         & \multicolumn{1}{c|}{}                       &                     & \multicolumn{1}{c|}{14}  & \multicolumn{1}{c|}{121.43}                    & \multicolumn{1}{c|}{}                         & \multicolumn{1}{c|}{}                       & \multicolumn{1}{c|}{1.32}               &                         & \multicolumn{1}{c|}{}                   & \multicolumn{1}{c|}{9}   & \multicolumn{1}{c|}{120.05}                     & \multicolumn{1}{c|}{}                         & \multicolumn{1}{c|}{}                       & \multicolumn{1}{c|}{1.85}        &                         \\ \cline{2-3} \cline{7-8} \cline{11-11} \cline{13-18}
                        & \multicolumn{1}{c|}{9}   & \multicolumn{1}{c|}{125}                   & \multicolumn{1}{c|}{}                         & \multicolumn{1}{c|}{}                       &                     & \multicolumn{1}{c|}{9}   & \multicolumn{1}{c|}{120.05}                    & \multicolumn{1}{c|}{}                         & \multicolumn{1}{c|}{}                       & \multicolumn{1}{c|}{1.85}               &                         & \multicolumn{1}{c|}{\multirow{4}{*}{3}} & \multicolumn{1}{c|}{6}   & \multicolumn{1}{c|}{58.91}                      & \multicolumn{1}{c|}{\multirow{4}{*}{5.85e-4}} & \multicolumn{1}{c|}{\multirow{4}{*}{0.694}} & \multicolumn{1}{c|}{3.03}        &                         \\ \cline{1-5} \cline{7-11} \cline{14-15} \cline{18-18}
\multirow{4}{*}{3}      & \multicolumn{1}{c|}{2}   & \multicolumn{1}{c|}{187.5}                 & \multicolumn{1}{c|}{\multirow{4}{*}{1.31e-3}} & \multicolumn{1}{c|}{\multirow{4}{*}{0.662}} &                     & \multicolumn{1}{c|}{2}   & \multicolumn{1}{c|}{178.1}                     & \multicolumn{1}{c|}{\multirow{4}{*}{7.92e-4}} & \multicolumn{1}{c|}{\multirow{4}{*}{0.681}} & \multicolumn{1}{c|}{2.37}               &                         & \multicolumn{1}{c|}{}                   & \multicolumn{1}{c|}{2}   & \multicolumn{1}{c|}{178.1}                      & \multicolumn{1}{c|}{}                         & \multicolumn{1}{c|}{}                       & \multicolumn{1}{c|}{2.37}        &                         \\ \cline{2-3} \cline{7-8} \cline{11-11} \cline{14-15} \cline{18-18}
                        & \multicolumn{1}{c|}{19}  & \multicolumn{1}{c|}{93.75}                 & \multicolumn{1}{c|}{}                         & \multicolumn{1}{c|}{}                       &                     & \multicolumn{1}{c|}{19}  & \multicolumn{1}{c|}{89.67}                     & \multicolumn{1}{c|}{}                         & \multicolumn{1}{c|}{}                       & \multicolumn{1}{c|}{1.02}               &                         & \multicolumn{1}{c|}{}                   & \multicolumn{1}{c|}{11}  & \multicolumn{1}{c|}{59.45}                      & \multicolumn{1}{c|}{}                         & \multicolumn{1}{c|}{}                       & \multicolumn{1}{c|}{1.04}        &                         \\ \cline{2-3} \cline{7-8} \cline{11-11} \cline{14-15} \cline{18-18}
                        & \multicolumn{1}{c|}{11}  & \multicolumn{1}{c|}{62.5}                  & \multicolumn{1}{c|}{}                         & \multicolumn{1}{c|}{}                       &                     & \multicolumn{1}{c|}{11}  & \multicolumn{1}{c|}{59.45}                     & \multicolumn{1}{c|}{}                         & \multicolumn{1}{c|}{}                       & \multicolumn{1}{c|}{1.04}               &                         & \multicolumn{1}{c|}{}                   & \multicolumn{1}{c|}{17}  & \multicolumn{1}{c|}{58.4}                       & \multicolumn{1}{c|}{}                         & \multicolumn{1}{c|}{}                       & \multicolumn{1}{c|}{1.42}        &                         \\ \cline{2-3} \cline{7-8} \cline{11-11} \cline{13-18}
                        & \multicolumn{1}{c|}{17}  & \multicolumn{1}{c|}{62.5}                  & \multicolumn{1}{c|}{}                         & \multicolumn{1}{c|}{}                       &                     & \multicolumn{1}{c|}{17}  & \multicolumn{1}{c|}{58.4}                      & \multicolumn{1}{c|}{}                         & \multicolumn{1}{c|}{}                       & \multicolumn{1}{c|}{1.42}               &                         & \multicolumn{1}{c|}{\multirow{3}{*}{4}} & \multicolumn{1}{c|}{18}  & \multicolumn{1}{c|}{57.54}                      & \multicolumn{1}{c|}{\multirow{3}{*}{6.81e-4}} & \multicolumn{1}{c|}{\multirow{3}{*}{0.691}} & \multicolumn{1}{c|}{3.54}        &                         \\ \cline{1-5} \cline{7-11} \cline{14-15} \cline{18-18}
\multirow{2}{*}{4}      & \multicolumn{1}{c|}{7}   & \multicolumn{1}{c|}{187.5}                 & \multicolumn{1}{c|}{\multirow{2}{*}{1.16e-3}} & \multicolumn{1}{c|}{\multirow{2}{*}{0.665}} &                     & \multicolumn{1}{c|}{7}   & \multicolumn{1}{c|}{177.43}                    & \multicolumn{1}{c|}{\multirow{2}{*}{8.54e-4}} & \multicolumn{1}{c|}{\multirow{2}{*}{0.685}} & \multicolumn{1}{c|}{2.55}               &                         & \multicolumn{1}{c|}{}                   & \multicolumn{1}{c|}{5}   & \multicolumn{1}{c|}{210.93}                     & \multicolumn{1}{c|}{}                         & \multicolumn{1}{c|}{}                       & \multicolumn{1}{c|}{1.67}        &                         \\ \cline{2-3} \cline{7-8} \cline{11-11} \cline{14-15} \cline{18-18}
                        & \multicolumn{1}{c|}{5}   & \multicolumn{1}{c|}{218.75}                & \multicolumn{1}{c|}{}                         & \multicolumn{1}{c|}{}                       &                     & \multicolumn{1}{c|}{5}   & \multicolumn{1}{c|}{210.93}                    & \multicolumn{1}{c|}{}                         & \multicolumn{1}{c|}{}                       & \multicolumn{1}{c|}{1.67}               &                         & \multicolumn{1}{c|}{}                   & \multicolumn{1}{c|}{19}  & \multicolumn{1}{c|}{89.11}                      & \multicolumn{1}{c|}{}                         & \multicolumn{1}{c|}{}                       & \multicolumn{1}{c|}{2.97}        &                         \\ \cline{1-5} \cline{7-11} \cline{13-18}
5                       & \multicolumn{1}{c|}{20}  & \multicolumn{1}{c|}{218.75}                & \multicolumn{1}{c|}{1.13e-3}                  & \multicolumn{1}{c|}{0.691}                  &                     & \multicolumn{1}{c|}{20}  & \multicolumn{1}{c|}{210.51}                    & \multicolumn{1}{c|}{8.33e-4}                  & \multicolumn{1}{c|}{0.715}                  & \multicolumn{1}{c|}{2.27}               &                         & \multicolumn{1}{c|}{5}                  & \multicolumn{1}{c|}{7}   & \multicolumn{1}{c|}{176.35}                     & \multicolumn{1}{c|}{5.83e-4}                  & \multicolumn{1}{c|}{0.728}                  & \multicolumn{1}{c|}{4.64}        &                         \\ \cline{1-5} \cline{7-11} \cline{13-18}
\multirow{3}{*}{6}      & \multicolumn{1}{c|}{3}   & \multicolumn{1}{c|}{187.5}                 & \multicolumn{1}{c|}{\multirow{3}{*}{1.14e-3}} & \multicolumn{1}{c|}{\multirow{3}{*}{0.661}} &                     & \multicolumn{1}{c|}{3}   & \multicolumn{1}{c|}{176.65}                    & \multicolumn{1}{c|}{\multirow{3}{*}{7.53e-4}} & \multicolumn{1}{c|}{\multirow{3}{*}{0.682}} & \multicolumn{1}{c|}{2.76}               &                         & \multicolumn{1}{c|}{\multirow{3}{*}{6}} & \multicolumn{1}{c|}{8}   & \multicolumn{1}{c|}{58.45}                      & \multicolumn{1}{c|}{\multirow{3}{*}{5.84e-4}} & \multicolumn{1}{c|}{\multirow{3}{*}{0.692}} & \multicolumn{1}{c|}{3.76}        &                         \\ \cline{2-3} \cline{7-8} \cline{11-11} \cline{14-15} \cline{18-18}
                        & \multicolumn{1}{c|}{18}  & \multicolumn{1}{c|}{62.5}                  & \multicolumn{1}{c|}{}                         & \multicolumn{1}{c|}{}                       &                     & \multicolumn{1}{c|}{18}  & \multicolumn{1}{c|}{57.95}                     & \multicolumn{1}{c|}{}                         & \multicolumn{1}{c|}{}                       & \multicolumn{1}{c|}{1.59}               &                         & \multicolumn{1}{c|}{}                   & \multicolumn{1}{c|}{1}   & \multicolumn{1}{c|}{89.45}                      & \multicolumn{1}{c|}{}                         & \multicolumn{1}{c|}{}                       & \multicolumn{1}{c|}{1.1}         &                         \\ \cline{2-3} \cline{7-8} \cline{11-11} \cline{14-15} \cline{18-18}
                        & \multicolumn{1}{c|}{1}   & \multicolumn{1}{c|}{93.75}                 & \multicolumn{1}{c|}{}                         & \multicolumn{1}{c|}{}                       &                     & \multicolumn{1}{c|}{1}   & \multicolumn{1}{c|}{89.45}                     & \multicolumn{1}{c|}{}                         & \multicolumn{1}{c|}{}                       & \multicolumn{1}{c|}{1.1}                &                         & \multicolumn{1}{c|}{}                   & \multicolumn{1}{c|}{3}   & \multicolumn{1}{c|}{176.65}                     & \multicolumn{1}{c|}{}                         & \multicolumn{1}{c|}{}                       & \multicolumn{1}{c|}{2.76}        &                         \\ \cline{1-5} \cline{7-11} \cline{13-18}
\multirow{3}{*}{7}      & \multicolumn{1}{c|}{8}   & \multicolumn{1}{c|}{62.5}                  & \multicolumn{1}{c|}{\multirow{3}{*}{1.19e-3}} & \multicolumn{1}{c|}{\multirow{3}{*}{0.659}} &                     & \multicolumn{1}{c|}{8}   & \multicolumn{1}{c|}{59.25}                     & \multicolumn{1}{c|}{\multirow{3}{*}{7.25e-4}} & \multicolumn{1}{c|}{\multirow{3}{*}{0.683}} & \multicolumn{1}{c|}{1.1}                &                         & \multicolumn{1}{c|}{\multirow{2}{*}{7}} & \multicolumn{1}{c|}{20}  & \multicolumn{1}{c|}{209.6}                      & \multicolumn{1}{c|}{\multirow{2}{*}{5.52e-4}} & \multicolumn{1}{c|}{\multirow{2}{*}{0.695}} & \multicolumn{1}{c|}{3.76}        &                         \\ \cline{2-3} \cline{7-8} \cline{11-11} \cline{14-15} \cline{18-18}
                        & \multicolumn{1}{c|}{4}   & \multicolumn{1}{c|}{187.5}                 & \multicolumn{1}{c|}{}                         & \multicolumn{1}{c|}{}                       &                     & \multicolumn{1}{c|}{4}   & \multicolumn{1}{c|}{181.83}                    & \multicolumn{1}{c|}{}                         & \multicolumn{1}{c|}{}                       & \multicolumn{1}{c|}{1.4}                &                         & \multicolumn{1}{c|}{}                   & \multicolumn{1}{c|}{10}  & \multicolumn{1}{c|}{58.52}                      & \multicolumn{1}{c|}{}                         & \multicolumn{1}{c|}{}                       & \multicolumn{1}{c|}{1.4}         &                         \\ \cline{2-3} \cline{7-8} \cline{11-11} \cline{13-18}
                        & \multicolumn{1}{c|}{10}  & \multicolumn{1}{c|}{62.5}                  & \multicolumn{1}{c|}{}                         & \multicolumn{1}{c|}{}                       &                     & \multicolumn{1}{c|}{10}  & \multicolumn{1}{c|}{58.52}                     & \multicolumn{1}{c|}{}                         & \multicolumn{1}{c|}{}                       & \multicolumn{1}{c|}{1.4}                &                         & \multicolumn{1}{c|}{\multirow{2}{*}{8}} & \multicolumn{1}{c|}{16}  & \multicolumn{1}{c|}{121.87}                     & \multicolumn{1}{c|}{\multirow{2}{*}{5.45e-4}} & \multicolumn{1}{c|}{\multirow{2}{*}{0.731}} & \multicolumn{1}{c|}{1.15}        &                         \\ \cline{1-5} \cline{7-11} \cline{14-15} \cline{18-18}
8                       & \multicolumn{1}{c|}{16}  & \multicolumn{1}{c|}{125}                   & \multicolumn{1}{c|}{1.01e-3}                  & \multicolumn{1}{c|}{0.71}                   &                     & \multicolumn{1}{c|}{16}  & \multicolumn{1}{c|}{121.87}                    & \multicolumn{1}{c|}{6.93e-4}                  & \multicolumn{1}{c|}{0.72}                   & \multicolumn{1}{c|}{1.15}               &                         & \multicolumn{1}{c|}{}                   & \multicolumn{1}{c|}{4}   & \multicolumn{1}{c|}{180.93}                     & \multicolumn{1}{c|}{}                         & \multicolumn{1}{c|}{}                       & \multicolumn{1}{c|}{3.43}        &                         \\ \hline
\end{tabular}
\end{table*}

\vspace{-0.1in}
\subsection{Negotiation via Incentivization}
We next evaluate the incentivization rules listed in Section \ref{sec_iv}), offered by the CPO. The results for each options are summarized in Table~\ref{sumtab4}, where $M^{\rm C\star}:=\sum M_v^{\rm C\star}$ and $M^{\star}:=\sum M_v^{\star}$. Due to page limit, we list the values of the charging currents only for a specific set of EVs. In option 1, each EV charges at their preferred time slot pairs $\{e_v^p,\tau_v^p\}$ (as shown in Fig~\ref{choice}) with the maximum charging rate $i^{\rm max}_v$. For instance, EVs 6, 12, and 15 in time slot 1 of case 1 requested charging rates of 62.5 A, 187.5 A, and 187.5 A, respectively. This rate yields the system $\mathcal{H}_2$-norm to be $1.09\times 10^{-3}$ and $\rm {VSI}_{\min} = 0.681$, which indicates a deteriorating system performance with \(M^{\rm C \star}=0\), as consumer preferences are fully prioritized. 
In option 2, the consumer preference is held fixed while charging the EVs at optimized charging rates. %For Case 1, the charging rates of EVs are optimized to 59.51 A, 181.38 A, and 182.43 A, respectively, which reduces the system $\mathcal{H}_2$-norm to $5.98 \times 10^{-4}$ from $1.09 \times 10^{-3}$, and improves ${\rm VSI}_{\rm min}$ to 0.709 from $0.681$. Thereafter, the CPO offers an incentive of \$1.02, \$1.52, and \$1.25 for engaging the EVs to option 2. Similarly, the charging rates of the remaining EVs across the eight slots are optimized, which costs a total incentive of \(M^{\rm C \star} = \$31.51\). 
Finally, in option 3, we prioritize grid stability over consumer preferences by reassigning EVs and adjusting their charging rates. For instance, in time slot 1 of case 1, option 3 suggests EV~6 and EV~12 to move to slots 3 and 2, and to charge at $58.91$ A and $180.92$ A, respectively, while asking EV~13 and 14 to move from time slot 2 to 1. This improves both the system \(\mathcal{H}_2\)-norm and the VSI to \(4.35 \times 10^{-4}\) and 0.718, respectively. We repeat this process across the eight time slots, which shows a total incentive of \(M^{\rm\star} = \$50.96\). The damping performance improves by 26.3\%-48.51\% with option 2 ($M^{\rm C \star} = \$31.51$) and by 20.25\% -39\% further with option 3 in exchange for an additional incentive of \$19.45. %The VSI also improves with option 2 and option 3 compared to option 1, as shown in Table~\ref{sumtab4}. Similar conclusions are seen to hold for case 2.  
The computation time for Case 1 is 195 seconds, as is similar for Case~2, which is compatible with day-ahead planning.
\vspace{-0.1in} 
\subsection{Closed-loop transient response}
\begin{figure}[h]
    \centering
    \includegraphics[width=0.44\textwidth]{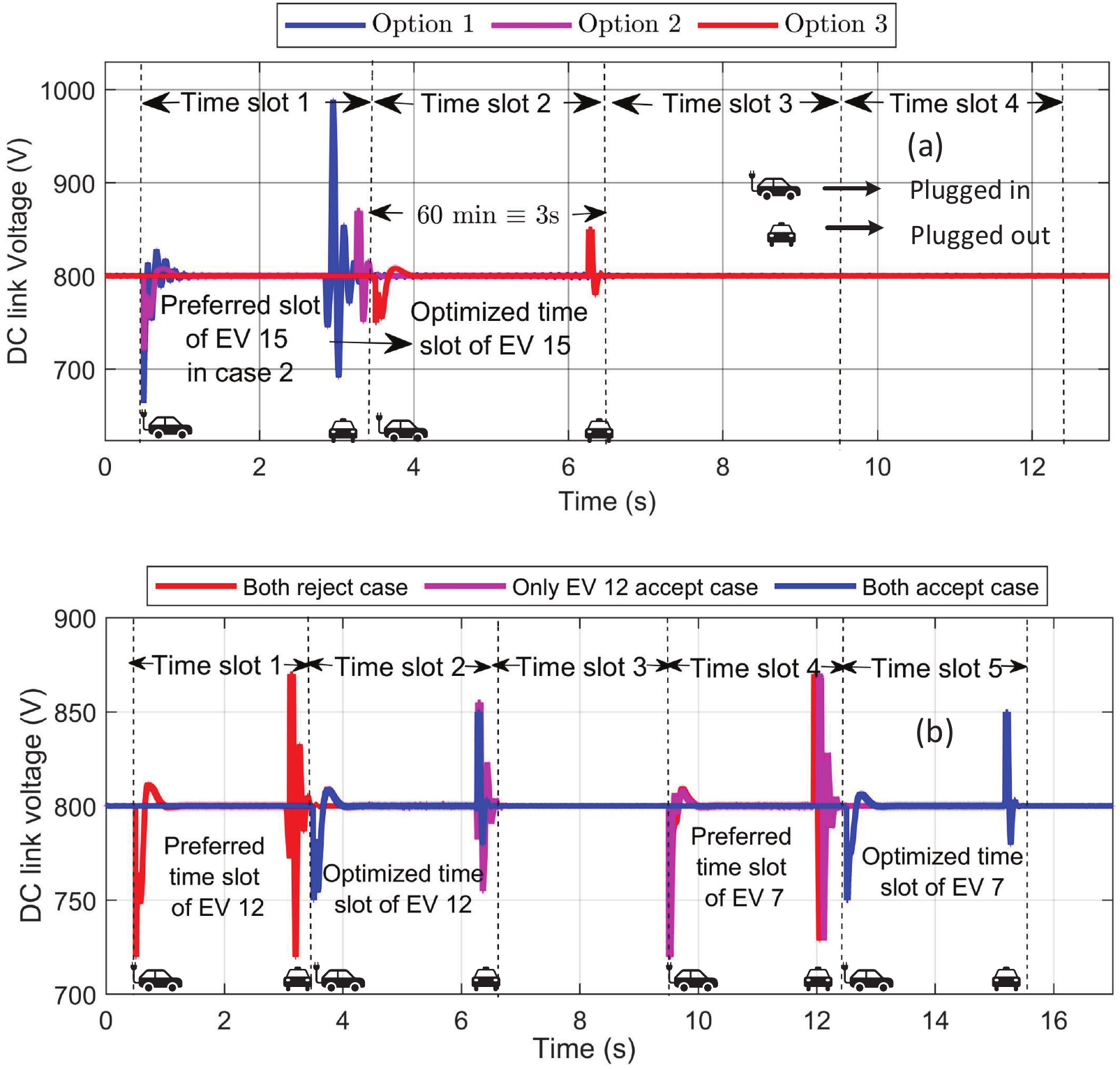}
    \caption{Transient DC link voltage profiles under representative EV charging scenarios: (a) comparison of system response under three options, (b) impact on system response based on EV consumer decision for EV charging.}
    \label{three_option_comparison}
\end{figure}

%We close this section by evaluating the transient characteristics of the power system in response to the plug-in and plug-out of the EVs that can be considered as a disturbance. 
Fig.~\ref{three_option_comparison}(a) compares the transient response of the DC link voltage at EVCS 1 in Case 2 during the plug-in and plug-out of EV~15 for the three charging options. The figure shows that choosing the optimized charging rate, even at the preferred time slot (which is option 2), substantially enhances the damping of the voltage response compared to option 1. Charging EV~15 at time slot 2 instead of its preferred time slot 1 with optimized charging rate (option 3) shows even better transient performance than option 2. Similarly, EVs 12 and 7 are intended to charge at time slots 1 and 4, respectively; however, the proposed optimized scheduling strategy suggests charging them at time slots 2 and 5, respectively. To demonstrate possible uncertainties in the customer behavior, we consider three types of responses from EV users under Case 2, namely: (i) both EVs reject the optimized scheduling and discount offer and choose option 1, (ii) only EV~12 accepts the offer in option 3, and (iii) both EVs accept the offer in option 3. The voltage profiles are shown in  Fig.~\ref{three_option_comparison} (b), which testifies that the damping of the DC link voltage oscillations is best when every driver accepts the CPO's offer.  % full acceptance of time slot coordination between EVs improves the damping of the voltage oscillations. These results collectively underscore the significance of leveraging the proposed approach for ensuring grid robustness while keeping the EV consumer convenience.
% \begin{figure}[!t]
%     \centering
%     \begin{subfigure}
%         \centering
%         \includegraphics[width=0.4\textwidth]{Fig_6a_v3_n.eps}
%         \label{three_option_comparison}
%         \vspace{-5pt}
%     \end{subfigure}
%     \hfill
%     \begin{subfigure}
%         \centering
%         \includegraphics[width=0.4\textwidth]{Fig_6b_v3.eps}
%         \label{accept_reject_case}
%         \vspace{-10pt}
%          % \caption*{(b)}
%     \end{subfigure}
%     \caption{Transient DC link voltage profiles under representative EV charging scenarios: (a) comparison of system response under three options, (b) impact on system response based on EV consumer decision for EV charging.}
%     \label{three_option_comparison}
% \end{figure}

% \begin{figure}[h]
%     \centering
%     \includegraphics[width=0.5\textwidth]{comparioson_figure.eps}
%     \caption{Accept reject case}
%     \label{accept_reject_case}
% \end{figure}

\section{Conclusions}
In this paper, we presented a two-stage optimization framework for scheduling the charging of large fleets of EVs in a power distribution grid. Unlike conventional scheduling methods that are usually based on steady-state objectives, our approach ensures improvement of small-signal stability, small-signal dynamic performance, and voltage stability of the distribution system model. This is enabled by an incentivization strategy that encourages EV drivers to shift their preferred time-slots for charging as well as waiting in the queue at the charging station for a few additional minutes in exchange for money. The framework can be especially useful for heavy-duty vehicles such as electric buses and trucks that have more flexibility in terms of charging time while driving on the highway during peak hours. Numerical results show that grid operators can benefit from this approach by enhancing their operating margins during heavy-load conditions. Our future work will extend this algorithm to flexible scheduling of other power-electronics-driven loads, such as AI data centers. 

\bibliographystyle{IEEEtran}
\bibliography{ref}
 \vspace{-50pt}
\begin{IEEEbiographynophoto}{Amit Kumer Podder}
received the B.S. and M.S. degrees in electrical engineering from Khulna University of Engineering \& Technology, Khulna, Bangladesh, in 2016 and 2019, respectively. He received a Ph.D. degree from North Carolina State University in Raleigh, USA, in 2025. His research interests include power systems dynamic stability analysis and control with high penetration of grid-interactive converters.
\end{IEEEbiographynophoto}
\vspace{-50pt}
 \begin{IEEEbiographynophoto}{Tomonori Sadamoto}
 received his Ph.D. degree from the Tokyo Institute of Technology, Tokyo, Japan, in 2015. From November 2018 to October 2023, he was an Assistant Professor with the Department of Mechanical and Intelligent Systems Engineering at the University of Electro-Communications. Since Nov. 2023, he has been an Associate Professor with the same department. In 2014, he was named a finalist for the 13th European Control Conference Best Student Paper Award. In 2020, he received the IEEE Control Systems Magazine Outstanding Paper Award. 
\end{IEEEbiographynophoto}
\vspace{-50pt}
 \begin{IEEEbiographynophoto}{Aranya Chakrabortty}
(F'25) received his PhD in Electrical Engineering from Rensselaer Polytechnic Institute in 2008. He is currently a Professor in the Electrical and Computer Engineering department at North Carolina State University, Raleigh, NC. His research interests are in all branches of control theory with applications to electric power systems. He received the NSF CAREER award in 2011 and was chosen as a Faculty Scholar by the NC State Provost's office in 2019.  
\end{IEEEbiographynophoto}
\end{document}